\documentstyle[lprocl,11pt]{article}
\input{psfig}
\def\be{\begin{equation}}
\def\bea{\begin{eqnarray}}
\def\ee{\end{equation}}
\def\eea{\end{eqnarray}}
\def\erf{{\rm erf}}

\begin{document}
\title{PROTEIN FOLDING AND HETEROPOLYMERS}
\author{T. GAREL, H. ORLAND, E. PITARD}
\address{Service de Physique Th\'eorique,\\
CEA-Saclay, 91191 Gif-sur-Yvette Cedex,\\ France}
\maketitle\abstracts{
We present a statistical mechanics approach to the protein folding
problem.
We first review some of the basic properties of proteins, and
introduce some physical models to describe their thermodynamics. These
models rely on a random heteropolymeric description of these non
random biomolecules. Various kinds of randomness are investigated, and
the connection with disordered systems is discussed. We conclude by a
brief study of the dynamics of proteins.}

 Natural proteins have the property of folding into an
(almost) unique compact native structure, which is of biological
interest \cite{Creighton84,Creighton92}. The compactness of this unique native state
is largely due to the existence of an optimal amount of
hydrophobic amino-acid residues \cite{Dill95}, since these biological objects are
usually designed to work in water. 
The task of predicting the conformation of the three-dimensional
structure from the linear primary sequence is often referred to as the
protein folding problem.
Already at this stage, a rigorous
analytical theory appears difficult, since it amounts to study a
mesoscopic system (most proteins have between 100 and 500 residues),
notwithstanding the solvent's properties.

This mesoscopic system is of a classical nature; the quantum
mechanical valence electrons of the atoms induce interactions between
the heavy ``nuclei'' which can then be treated as classical
objects interacting through classical many-body interactions.
This observation forms the basis of Molecular Dynamics, Monte Carlo,
or Statistical Mechanical models.

To further complicate the matter, 
both the compactness and the chemical heterogeneity of a
given protein tend to slow down dynamical processes: the question of
the kinetic control of the folding (as opposed to a thermodynamic
control) is therefore periodically asked. This remark suggests that
the folding process may have something in common with the physics of
glassy systems, where competing interactions (frustration) and/or
disorder lead to a very rugged phase space, resulting in slow dynamical
processes. In this review, we will assume that the folded state of
proteins is thermodynamically stable (the folded state is the state of
minimal free energy).

There has been a considerable amount of numerical simulations of
proteins, using molecular dynamics or Monte Carlo calculations (for a
review see \cite{Creighton92,Dill95,Elber96}).This promising
approach will not be discussed in this review. We will use only
analytical approaches throughout.
Similarly, models
emphasizing the micro-crystalline character of the folded proteins will
not be addressed here \cite{Garel94}.

The outline of this review is the following.
In the first section, we introduce at an elementary level,
some notions of the physics, chemistry and biology of proteins.
In the second section, we study (using statistical physics methods) 
some heteropolymer models which are
possibly relevant to the protein folding problem. The phase diagrams of
these models bear
some qualitative resemblance to the real systems. 

In the last section,
we tackle the issues of dynamics. In view of the
complexity of the problem, we first study the homopolymer case. The
heteropolymeric aspect is then studied at a more phenomenological level.

\section{Biophysical background}
For a review on the biological aspects of protein, the reader is referred to 
the books \cite{Dar_Lod_Bal,Stryer,cell}.
\subsection{Elements of Chemistry}
Proteins are biological molecules, present in any living
organism. Their biological function include catalysis (enzymes),
transport of ions (hemoglobin, chlorophyll, etc...), muscle contraction, ...
They also are present
in virus shells, prions, etc. The biological
activity involves a small set of atoms, called the ``active
site'' of the protein, where chemical reactions take place.

Proteins belong to the group of biopolymers, which also
comprise nucleic acids (DNA, RNA) and polysaccharides.
From a physico-chemical point of view, biopolymers are heteropolymers,
made out of different species of monomers.  For proteins, the monomers are
amino-acids, chosen from twenty different species. 

The chemical formula of an amino-acid can be written as:
\be
\label{aa}
NH_2 -C_{\alpha}HR-COOH
\ee
where $NH_2$ is the amine group and $COOH$ is the acidic group
(except for proline, which has an imine group).
Each amino acid is characterized by its residue $R$ . If the residue is
not reduced to a hydrogen atom (such as {\it glycine}), the
alpha-carbon atom $C_\alpha$
is asymmetric. In all known natural proteins,
the alpha-carbons have the same chirality (they are all left-handed)
and the origin of this asymmetry is not known.

The list of amino acids is:
\begin{enumerate}
\item \label{en1} Alanine, Isoleucine, Leucine, Methionine, Phenylalanine, Proline,
Tryptophan, Valine.
\item \label{en2} Asparagine, Cysteine, Glutamine, Glycine, Serine, Threonine,
Tyrosine.
\item \label{en3} Arginine, Histidine, Lysine.
\item \label{en4} Aspartic acid, Glutamic acid.
\end{enumerate}

For instance, the chemical formula of alanine is:
\be
NH_2 -C_{\alpha}H-CH_3-COOH
\ee
and that of tryptophan is:
\be
NH_2 -C_{\alpha}H-CH_2-C-CH-NH-C_6H_4-COOH
\ee

The smallest residue is glycine, which is just a single
$H$ atom, and thus non chiral, and the largest is tryptophan, which
contains ten heavy (non hydrogen) atoms.

The above gross classification of amino acids refers to their interactions
with water, their natural solvent. Group \ref{en1} is made of non
polar hydrophobic residues. The three other groups are made of
hydrophilic residues. From an electrostatic point of view, group
\ref{en2} corresponds to polar neutral residues, group \ref{en3}
corresponds to positively charged residues, and group \ref{en4} to
negatively charged ones.
       
The typical size of a protein ranges from approximately 100 amino acids
for small proteins  to 500 for long immuno-globulins. Due to this
rather small size, knots are not present in proteins.

A protein is made by polycondensation of amino acids, which can be
schematically written as:
\bea
\label{pc}
&&NH_2-C_{\alpha}HR_1-COOH + NH_2-C_{\alpha}HR_2-COOH  \to
\nonumber \\
&&NH_2-C_{\alpha}HR_1-CONH -C_{\alpha}HR_2-COOH + H_2O
\eea
The repetition of this process produces the protein, a weakly branched
polymer, characterized by its chemical sequence $R_1, R_2, \cdots,
R_N$. 

\begin{figure}
\centerline{
\psfig{figure=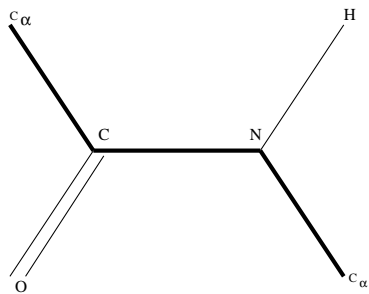,height=1.5in} }
\caption{Peptide bond. The thick line denotes the backbone.
\label{bond}}
\end{figure}

The polycondensation produces a ``peptide bond'' $CONH$, represented in
Fig.~\ref{bond}.

 Due to electronic hybridization, this bond is strongly planar.

One can distinguish two types of degrees of freedom in proteins:
\begin{enumerate} 
\item Hard degrees of freedom: these are the covalent bonds
(linking covalently two atoms along the chain), the valence angles
(angle between two covalent bonds) and the peptide bond. These degrees
of freedom are very rigid at room temperature, since, as we shall see
later, their deformation requires energies much higher than $kT$.
\item Soft degrees of freedom: they are essentially the torsion angles
along the backbone chain, and of the side chains. Their energy scale is such
that they can easily fluctuate at room temperature.
\end{enumerate}

\subsection{The possible states of proteins}

\subsubsection{Qualitative description of the phases}

Although it was long believed that proteins are either denatured or native,
it seems now well established that they may in fact exist in at least three different phases.
Originally, the phases referred to the biological activity of the
protein. In the ``native phase'', the protein has its full biological
activity, whereas in the ``denatured phase'', it does not have any
biological activity. It was soon recognized that this change in
activity was related to some important structural changes in the
protein. 

The following classification is widely accepted:
\begin{enumerate}
\item {\bf Native state} \\
In this phase, the protein  is said to be folded and has its full biological activity; it is compact,
globular, and has a unique and well defined three dimensional
structure. This implies that in this state,
the active site is well defined.

As we shall discuss later, the uniqueness of the folded state is quite
puzzling, given that the number of compact states of a homopolymer of size
$N$ is known to behave like $\mu^N$ \cite{descl}. Even with an extremely
conservative value $\mu=2$ and $N=100$, this represents an astronomically
large number of compact configurations, and it is quite amazing that
proteins always find their ways to the correct folded
state. Basically, the conformational entropy of the native state is zero.

In Fig.~\ref{shape}, we show a graphical representation of the protein
crambin, generated from its X-ray data.
\begin{figure}
\centerline{
\psfig{figure=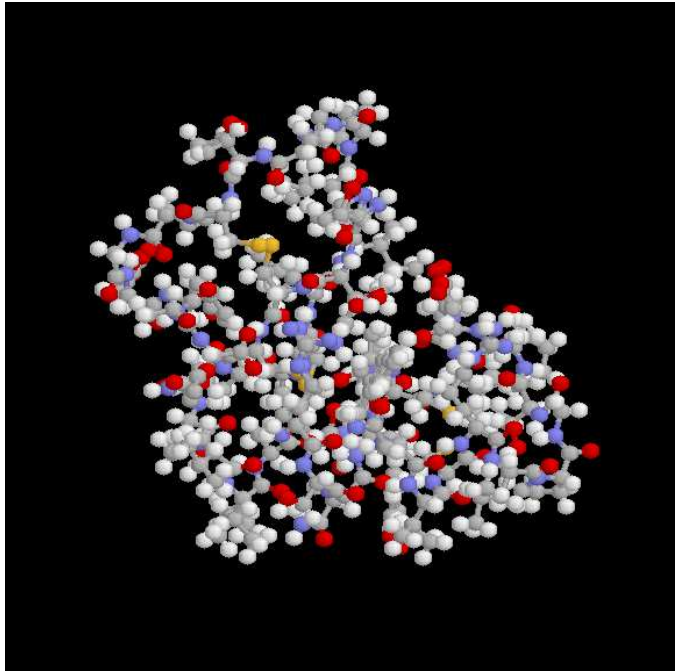,height=3.5in} }
\caption{Computer generated ``ball and stick'' representation of crambin.
\label{shape}}
\end{figure}

\item {\bf Denatured states} \\
The denatured states are characterized by a lack of biological activity
of the protein. Depending on chemical conditions, it seems that there
exists at least two denatured phases:
\begin{enumerate}
\item {\it Coil state} \\
In this state, the denatured protein is in a coil state: it has no
definite shape, like a homopolymer in a good solvent. Although in this
phase there might be local aggregation phenomena, it is fairly well
described as the swollen phase of a homopolymer in a good solvent.
It is a state of large conformational entropy.

\item {\it Molten globule} \\
At low pH (acidic conditions),  some proteins may exist in a compact
state, named ``molten globule'' \cite{Pti}. This state is compact (the chain is
globular); it does not however have a well defined structure and bears
strong resemblance to the collapsed phase of a homopolymer in a 
bad solvent. It is believed that this state is slightly less compact
than the native state, and has finite conformational
entropy. Anticipating on the next sections, the molten globule seems
to have a large content of secondary structure, but not necessarily at
the right place (with respect to the native state).
\end{enumerate}
\end{enumerate}

In vitro, the transition between the various phases is controlled by
temperature, pH, denaturant agent (such as urea or
guanidine). 

\subsubsection{Time scales}

There are two basic time scales in the problem:
\begin{enumerate}
\item Microscopic \\
The shortest time involved in protein dynamics is related to the
vibrational modes of the covalent bonds. The associated time scale is
$10^{-15}$s.
\item Macroscopic \\
Typical times for the folding of a protein ranges from $10^{-3}$s to
$1$s. This is many order of magnitudes larger than the microscopic
time, and it is quite puzzling to understand why
such a long time is necessary for such a small system to relax to equilibrium.
As we shall see, the main reason is that the energy landscape of a
compact chain is very rugged, with an exponentially large number of
metastable states, separated by high barriers. This situation is
familiar in spin-glasses or other disordered systems. We already
mentioned
that even in
a homopolymer chain, there is of the order of $\mu^N$
compact quasi-degenerate ground states, separated by high barriers. 
\end{enumerate}

\subsubsection{Experimental techniques}
There are several techniques which allow one  to study the folded
structure of proteins,
and each one gives access to a different aspect of the
problem.

\begin{enumerate} 
\item {\it Biological techniques}: they mainly use the
recovery of the biological activity of the proteins. These
techniques allow measurement of rate constants, reaction yields, etc...
\item{\it Measurements of the radius of gyration}: one can measure the
radius of gyration, as well as the structure factor of
proteins, by use of X-ray or neutron scattering.
\item{\it NMR}: this technique allows to detect neighboring pairs of
resonating protons which in turn give strong constraints for  the 
spatial resolution of
structures.
\item{\it Circular dichro\"\i sm}: CD allows to look for
secondary structures (which will be defined slightly later). It is
sensitive to optical activity (due to the presence of $\alpha$-helices).
\item{\it X-ray crystallography}: this is the most precise
method to
solve the three dimensional structure of proteins. In order to get
any structural information about proteins, 
it is necessary to crystallize the proteins so as to freeze
the positions of the atoms. This is a very difficult task,
since one must first make a crystal from the
protein, and then resolve its structure (in particular, one has to
resolve the phase ambiguity).
\end{enumerate}
\subsection{The different structural levels}
Since the discovery and resolution of many protein structures, it has
become customary to distinguish several levels of organization in the
structure.
\subsubsection{Primary structure}
The primary structure is just the chemical sequence of amino acids
along the main
backbone chain. This chemical structure is routinely
determined experimentally, by using techniques such as electrophoresis,
etc...
\subsubsection{Secondary structures}
Pauling  and Corey\cite{pau_co} first predicted theoretically that proteins should
exhibit some local ordering, now known as secondary structures. Their
prediction was based on energy considerations: they showed that 
there are certain
regular structures which maximize the number of Hydrogen bonds (H-bonds)
between the C-O  and the H-N groups of the backbone. There are
basically two such types of structures:
\begin{enumerate}
\item {$\alpha$-helices} \\
These are one-dimensional structures. The H-bonds are aligned with the
axis of the helix (see Fig.~\ref{helix}). There are 3.6 amino acids per helix
turn, and the typical size of a helix is 5 turns.

\begin{figure}
\centerline{
\psfig{figure=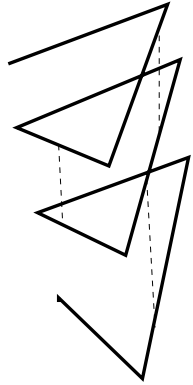,height=2.0in}}
\caption{A $\alpha$-helix. The dashed lines represent the H-bonds.
\label{helix}}
\end{figure}

\item {$\beta$-sheets} \\
	These are quasi two-dimensional structures. The H-bonds are
perpendicular to the strands. A typical $\beta$-sheet has a length of
8 amino acids, and consists of approximately 3 strands  (see
Fig.~\ref{beta}).

\begin{figure}
\centerline{
\psfig{figure=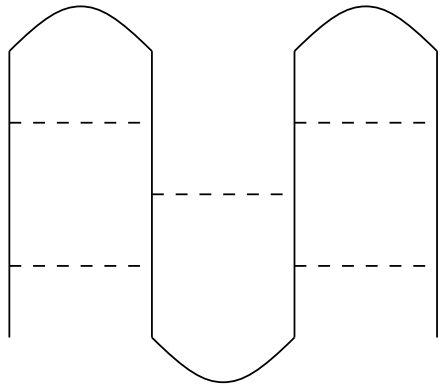,height=2.0in}}
\caption{A $\beta$-sheet. The dashed lines represent the H-bonds.
\label{beta}}
\end{figure}

\end{enumerate}
\subsubsection{Tertiary structure}
It is the compact packing of the secondary structures which
make up the tertiary structure: it is essentially the full three
dimensional structure of the protein.
\subsubsection{Quaternary structure}
Large proteins are made up of entities called domains, which are
compact globular regions, separated by a few amino acids. These domains
are mobile one relative to the others and their arrangement  is called the
quaternary structure.

\subsection{Interactions}
It is important to analyze the interactions between the various atoms
present in the system. At the microscopic level
there is only Coulombic interactions; such a microscopic approach is at present out of reach.

Chemists and physicists have instead analyzed interactions 
at a more macroscopic level:
they introduce semi-empirical interactions which can then be
included in energy minimizations, molecular dynamics, Monte Carlo
calculations, etc...

Along these lines, one may distinguish two main types of interactions:
\begin{enumerate}
\item{\it bonded} \\
These are the covalent bonds between the atoms of the protein. One may
further define:
\begin{enumerate}
\item{connectivity bonds}: they are the chemical bonds along the backbone or
side chains.
\item{sulfur bridges}: these are covalent bonds which may form 
only between the sulfur atoms of the cysteine residues.
\end{enumerate}
\item{\it non bonded} \\
There are several types of non-bonded interactions in proteins:
\begin{enumerate}
\item{Coulomb}: some atoms are assigned  partial charges (smaller than one
electronic charge), and interact through Coulomb
interaction (the question of the relative dielectric constant is a matter of
debate).
\item{Van der Waals}: this interaction accounts for the strong
steric repulsion at short distances, and the long range dipolar attraction at
larger distances. It is usually represented by a Lennard-Jones 6-12
potential of the form:
\be
v({\vec r}) = {A \over r^{12}} -  {B \over r^{6}}
\ee
\item{Hydrogen bonds}: the interaction responsible for the formation
of H-bonds can be introduced explicitly by a 6-10 potential similar
to the Lennard-Jones potential, but it is now quite accepted that H-bonds are just
a result of the combination of Coulomb and Van der Waals interactions.
\end{enumerate}
\item{The role of water}: water is a dipolar molecule, and thus has
strong interactions with charged or dipolar groups (see the above
classification of the residues). Since proteins are active in an aqueous
environment, water must be taken into account. This is the origin of the
hydrophobic effect: hydrophobic groups will be buried inside the
globule, whereas hydrophilic groups will be on the surface, in contact
with water. However, when one looks at the database of real proteins,
it turns out that the situation is not so clear cut: there is a
substantial probability ($\sim 35 \%$) 
to find a hydrophobic residue on the surface
of a protein, or to find a hydrophilic one buried inside.  The
situation is 
simpler for
charged residues (including the ends of the chain which are ionized in
water), which are almost always
found on the outer surface of the protein.
\end{enumerate}

\subsection{Energy scales}
The natural energy scale in protein chemistry is the kilo Calorie
per mole, denoted kCal/mole. The correspondence with more physical
units is $300 K \simeq 0.6 \ $kCal/mole. 

The typical denaturation temperature for a protein is 1 kCal/mole.

There are two widely
separated energy scales involved:
\begin{enumerate}
\item bonded interactions: their energy range is from 50 kCal/mole to 150
kCal/mole. They correspond typically to $100\ k_BT$ at room
temperature, and therefore, are not excited thermally.
\item non bonded interactions: their energy range is from $1$ to 5
kCal/mole. They are thermally excited at room temperature, and are
thus responsible for the folding and all the observed thermodynamical
properties of proteins.
\end{enumerate}
In a simplified description, it seems natural to consider all the
bonded interactions as frozen
(implying that the primary structure is quenched), 
and take into account only the non
bonded terms.

\subsection{Summary}
To summarize this section, we emphasize again that the
hydrophobic
interaction is a strong driving force for the collapse of proteins. In
addition, there are competing interactions between the various
amino acids (Coulomb, Van der Waals). This energetic frustration,
together with the topological constraints induced by the chain, give
rise to the existence of an exponential number of metastable states.
The folded state
is a compromise which minimizes the total free energy of the protein,
subject to the chain constraint.
In the following chapters, we shall present schematic heteropolymer
models to describe some aspects of the physics of proteins. The
questions we will more specifically address are:
\begin {enumerate} 
\item What is the nature of the transition: is it a
liquid to crystal type of transition, or rather a glass transition?
\item How can one understand the unicity of the folded state, given
the extremely large number of metastable states.
\item What are the mechanisms of folding? How can one describe the
short and  long time dynamics of proteins.
\end{enumerate}

\section{Heteropolymer models for proteins}
\subsection{Introduction}
\label{sec:intro}
As mentioned above, there are two main approaches to
the (in vitro) protein folding transition. One may first view the
folded state as an ordered microcrystal, with helix-like and
sheet-like domains. This type of approach clearly relies on the
universal character of secondary structures in proteins, and will not
be considered here. 
\par
In this review, we  will be concerned with the tertiary structure, and
rather emphasize the heterogeneity of the primary
sequences of proteins. These non-random (evolution-selected)
macromolecules will be modelled by disordered polymers of various
sorts. In these models, the 
relevant interactions (monomer-monomer or monomer-solvent) are
disordered, and the disorder is quenched, to account for the fixed
character of the chemical sequence $({R_1,R_2,...R_N})$. 
This hypothesis of a
quenched disorder is not always obvious (there may be for instance
an electrostatic charge on a monomer which depends on the monomer
environment, hence an annealed character) but we will restrict our
study to this case. This quenched disorder approach has clearly a lot
in common with spin glass transitions. In particular it is tempting to
consider the native state of an heteropolymer as the ``almost unique''
frozen state of the type present in spin glass
mean field theories. A major difference though with spin glass theories is
linked to the chain constraint, which, inter alia, induces long-range
interactions along the primary sequence. 

The problems one faces in heteropolymers are therefore twofold,
concerning both the ``hetero'' and the ``polymer'' aspects. 
As in spin glasses \cite{Me_Pa_Vi}, it is not easy to have
precise informations for a fixed disorder configuration (i.e. a single
chain); one is often
led to average over all possible disorder configurations
(i.e. over all possible disordered chains). One then follows ``well
established'' replica routes (mean-field, variational
approach,....). It is a characteristic feature of spin glasses that
this thermodynamic approach often leads to metastable states, and
therefore raises questions on the dynamics of the system.\par
The frequent inadequacy of the heteropolymer  
approach to deal with a fixed primary sequence makes the connexion
with biology rather tenuous and does not help much in bringing
biologists and physicists together. Furthermore, the thermodynamic limit that
one considers in heteropolymer studies must be cautiously applied to a
100 monomers protein, not to mention the dynamical problems previously
alluded to.
If folding transition is understood as ``strongly cooperative folding
process'', we nevertheless believe that such 
an approach is useful since it may connect the unusual folding transition
with the unusual transitions of disordered condensed matter physics. 
We have adopted the following plan to disentangle as much as possible
the difficulties linked to the two aspects of heteropolymers. Most of
the basic polymer theory has been postponed to the Appendices. A brief
summary of spin glass thermodynamics is presented in section
\ref{sec:Summary}. Section \ref{sec:Models} deals with various
models of randomness that one may find in
heteropolymers, and their physical interpretation. The random 
hydrophilic-hydrophobic chain will be considered in section \ref{sec:hydro}. In
section \ref{sec:ranbon}, we consider the ``random bond'' model which has a Random
Energy Model type freezing transition in high dimensions. Other types
of randomness are briefly considered in section \ref{sec:ranseq}. The relevance of
these ideas for real proteins is briefly examined in subsection \ref{sec:hetero}. 
Let us stress here the fact that the state of the art
concerning the three dimensional heteropolymer folding problem is not
even comparable to that of spin glasses.  

\subsection{A brief summary of spin glass theory}
\label{sec:Summary}
A detailed state of this field is treated elsewhere in this book
\cite{book}, so we will content ourselves with the minimum amount
needed for the heteropolymer folding problem. The spin glass problem
originally arose in the study of disordered magnetic alloys, where magnetic
impurities (e.g. Mn), quenched at random positions in the host lattice
(e.g. Cu) interact with one another through competing (oscillating)
interactions. This model has been largely extended
\cite{Me_Pa_Vi,Bin_You}, and the interactions between the spins are
accordingly of various sorts (exchange, dipolar,
Dzialoshinsky-Morya,...). There is however a general consensus that 
the very essence of the spin glass problem (i.e. frustration + quenched
disorder) is captured by the Edwards-Anderson Hamiltonian:
\be
\label{hamea}
{\cal H} = -\sum_{i<j} J_{ij} \ {S}_i {S}_j
\ee
where ${S}_i=\pm1$ is the (Ising) spin of impurity $i$ at position ${\vec
r}_i$ and $J_{ij} = J({\vec r}_i,{\vec r}_j)$ is the random coupling
between impurities $i$ and $j$, distributed with a law $P(\{J_{ij}\})$. 
The frustration stems from the fact that the
couplings $\{J_{ij}\}$ may take positive (ferromagnetic) or
negative (antiferromagnetic) values. Furthermore, the position $\vec
r_i$ of the impurities (and therefore the $J_{ij}$ themselves) are
quenched variables, which means that they evolve with a time scale
infinitely longer than the time scales of the thermodynamic variables
$S_i$ . This implies that for a given set of $\{J_{ij}\}$, the
free energy $F(\{J_{ij}\})$ at temperature $T$ is obtained as:
\be
\label{diso}
F(\{J_{ij}\}) = -T \log Z(\{J_{ij}\})
\ee
with
\be
\label{parti}
Z(\{J_{ij}\}) = {\rm Tr}_{{\{S_i}\}} \exp \left( -\beta {\cal H}(\{J_{ij}\})\right)
\ee
with $\beta ={1 \over T}$ (throughout this article, we set Boltzmann's
constant $k_B=1$). Note that $F(\{J_{ij}\})$ is also a random
variable; for short range 
interactions, we now show that, when the number $N$ of spins becomes 
very large,
 $F(\{J_{ij}\})$ is sharply peaked around its (disorder) averaged
value $\overline{ F}$, where
\be
\label{moy}
\overline{F} = \int \prod P(\{J_{ij}\})
F(\{J_{ij}\}) d(\{J_{ij}\})  
\ee
This property is called self-averageness of the free energy, and
the argument goes as follows \cite{Brout}: suppose we divide a $d$-dimensional system of $N$
spins $S_i$ into subsystems containing each $m$ spins with $1 <<
m << N$. The system's total free energy is the sum of 
two contributions:

(i) a contribution from each subsystem, of order $m$ per
"domain". \par
ii) a contribution from each "domain wall" between
neighboring subsystems, of order $m^{{d-1\over d}}$ for short-range
interactions. 

For $m$ large, the second contribution is negligeable compared to the
first. Equation (\ref {moy}) results by considering each subsystem to
represent a realization of the couplings $\{J_{ij}\}$. If $l$ denotes
a coherence length of the system, $m$ can be thought of as the number
of spins in a coherence volume $l^d$ (for a recent discussion, see
\cite{Aha_Har} and references therein). 

It is fair to say that the most
detailed study of spin glasses has been made with great difficulty in
the mean-field limit (infinite range interactions). In this case, it
can also be shown that the self-averageness of the free-energy holds in
the thermodynamic limit. The original Edwards-Anderson Hamiltonian has
been widely generalized (separable interactions, vectorial spins, Potts
variables,...). Restricting the discussion to Ising spins, 
we consider a typical spin glass Hamiltonian to be given by:
\be
\label{hamp}
{\cal H}_p = -\sum_{1\le i_1<i_2<\cdots<i_p\le N} 
J_{i_1 \cdots i_p} \  S_{i_1} \cdots S_{i_p}
\ee
where $p\ge 2$, and the couplings $\{J_{i_1 \cdots i_p}\}$ are
independent random 
variables, distributed according to a Gaussian law: 
\be
\label{proba}
h(J_{i_1 \cdots i_p})= \sqrt{{N^{p-1} \over \pi \ p!}}
\exp \left( {-{N^{p-1} \over p!} {J_{i_1 \cdots i_p}^2 \over J^2}}\right)
\ee
In (\ref {proba}), the $N$-dependent normalization ensures that the free
energy is extensive. 
\par
 Two main lines have been pursued in the (static) study of mean field
spin glasses \cite{Me_Pa_Vi,Bin_You}: 

1) the replica method where one averages a priori over all disorder
configurations, through the identity: 
\be
\label{log}
\overline{F}= -T \overline{\log Z} = -T \lim_{n \to 0} {
\overline{Z^n} -1 \over n}  
\ee

2) the TAP equations, which give, for each disorder configuration 
$\{ J_{i_1 \cdots i_p} \}$ the free energy local minima. These
equations are difficult to study either analytically or numerically,
except in certain limits. 
 
Both methods point towards a broad division of spin glass mean field
models into two categories

(i) usual spin glasses with a continuous Parisi replica symmetry
breaking scheme ($p=2$). The high temperature
phase has only the paramagnetic solution, whereas the low
temperature phase has an exponentially large number of TAP solutions,
among which the low lying free energy solutions build up the Parisi
replica order parameter.
 
(ii) spin glasses with a one step replica symmetry breaking scheme
($p \ge 3$). These models have a low temperature phase rather similar to
the previous ones, but possess two disordered phases above the
static critical temperature: one with an exponential number of
TAP metastable states, separated by high energy barriers, and the regular
paramagnet with no metastability at all. These metastable states have
of course dynamical significance (that is why these models may be
close to the real glass transition \cite{Ki_Thi_Wo}). Moreover, since
we are dealing with mean field models, they have (in the thermodynamic
limit) an  infinite lifetime, and must therefore be included in the
thermodynamics. 

The difference between these two classes may be traced back
to their behavior in the absence of disorder: when pure, the latter
models undergo first order transitions, implying the existence of a
spinodal (dynamical) temperature above the critical temperature.  \par  

It is important to note that the $p \to \infty$ model can be
identified with Derrida's Random Energy Model (REM) \cite{Der}, which
appears as a generic model in the physics of mean field disordered
systems. For instance, if one 
replace the Ising spins of the Edwards-Anderson model by $q$-state
Potts variables, one also \cite{Gro_Som} recovers the REM in the limit $q \to
\infty$.

Beyond the mean-field picture, one has basically to resort to Imry-Ma
like domain arguments \cite{IM,Berk}, to variational methods or to numerical
calculations. In particular, the very
existence of a spin glass phase in three dimensions, not to mention
its nature, is still an unsettled question. Related models of interest
\cite{book} include the random field Ising model, the role of
impurities on the Abrikosov vortex lattice,..., and the heteropolymer
folding transitions that we now present. 

\subsection{Quenched disorder in polymers}
\label{sec:diso}
In the case of heteropolymers, we are
interested in the statistical mechanics of a $d$ dimensional polymer
chain, with random quenched interactions (either with the solvent or with
itself). The positions of monomer $i, \ (i=1,2,...N)$ is denoted by $\vec
r_{i}$. The frustration in this case stems from conflicting terms in the
Hamiltonian and from the geometric chain constraint $g(\vec
r_{i}, \vec r_{i+1})$.
Throughout this work, we will restrict ourselves to the simplest forms
of chain constraint, namely
\par
(i) $ g({\vec r_{i}, \vec r_{i+1}})=\delta(\vert\vec r_{i} - {\vec r_{i+1}}\vert- a)$ for discrete chains ($a$ is the monomer length)\par
(ii) $g(\vec r_{i}, \vec r_{i+1}) \to \exp ( - {d \over 2 a^2 }\left(
{d {\vec r(s)} \over ds} \right)^2) $ for a continuous description of
the chain ($s$ denoting the curvilinear abscissa along the chain).\par
Other choices  are discussed in appendix A. Furthermore, we will
only  consider randomness in the two body interactions $v_{ij}(\vec
r_i,\vec r_j)$ or $v_{s,s'}(\vec r_s,\vec r_{s'})$. Whenever
necessary, we will also include the usual (homopolymer) many body
interactions (e.g. $ w_0(\vec r_i,\vec r_j,\vec r_k)$ for the three body
term). 
Choosing the discrete notation, the partition function of the
heteropolymer chain reads 
\begin{equation}
\label{part}
Z(\{v_{ij}\}) = \int \prod_{i}  d{\vec r_i} \prod_{i} 
g({\vec r_i},{\vec r_{i+1}}) \ \exp ( - \beta {\cal H}(\{v_{ij}\}) )
\end{equation}
where the (reduced) Hamiltonian is
\be
\label{hami}
\beta {\cal H}(\{v_{ij}\})={1 \over 2} \sum_{i\ne j}v_{ij}(\vec
r_i,\vec r_j)+ {1 \over 6} \sum_{i\ne j\ne k}w_0(\vec r_i,\vec
r_j,\vec r_k)+...
\ee	
and the dots $...$ include the possibility of higher order terms.
Its free energy reads
\be
\label{dis}
F(\{v_{ij}\}) = -T \log Z(\{v_{ij}\})
\ee
The self-averageness argument of the free energy may be presented in a
slightly different way from the spin glass case. Consider a ``soup''
of $M$ random chains of $N$ monomers (a monomer should be thought of
as an amino acid). Each of these chains represents
a different choice of $\{v_{ij}\}$ , i.e. a different
primary sequence. The free energy per chain of the soup is given by: 
\begin{equation}
\label{free1}
F = {1 \over M} \sum_1^M  F(\{v_{ij}\})
\end{equation}

This expression neglects the interchain interactions, and is thus
a priori valid, provided that (i) the interactions in the soup
are short-ranged (ii) the soup is dilute enough (one chain
problem), or concentrated enough (in a melt, interactions between the
chains are screened). \par
For large $M$, one may interpret $F$ as an average over all possible
choices of $\{v_{ij}\}$. Denoting this average by $ \overline{F}$, we have:
\begin{equation}
\label{free2}
F= \overline{F} = \int \prod P(\{v_{ij}\})
F(\{v_{ij}\}) d(\{v_{ij}\})  
\end{equation}
For a single chain (dilute soup problem), this point of view clearly
leads to the use of replicas, in order to 
perform the quenched average. This would yield the typical properties
of a typical chain of the soup. In a melt (concentrated soup problem),
one may avoid the use of replicas \cite{Fr_Mi_Le}, by using equation
(\ref{free1}). On 
the contrary, if one wants to study a specific primary sequence,
i.e. a given set of $\{v_{ij}\}$ (without any averaging procedure),
one must resort to the  self-consistent field method, described in
appendix B. Equations (\ref {ed7}) are in some sense the TAP
equations of our problem. Solving these equations would require an
involved numerical treatment, which has not been undertaken so far.\par

We now examine various possible choices of the two-body interaction term
$v_{ij}(\vec r_i,\vec r_j)$, with the (reasonable) restriction that we consider
only translationally invariant forms.\par
\be
\label{free2bis}
v_{ij}(\vec r_i,\vec r_j)=v_{ij}(\vec r_i-\vec r_j)
\ee
  One may first consider each monomer ${i}$
to be characterized by a single random (scalar) ``charge'' ${\xi_i}$: the
most general choice \cite{Obu} then reads
\be
\label{free3}
v_{ij}(\vec r_i-\vec r_j)=v_0(\vec r_i-\vec r_j)+\beta a(\vec r_i-\vec
r_j)(\xi_i+\xi_j)+\beta b(\vec r_i-\vec r_j)\xi_i\xi_j
\ee
where $v_0(\vec x), a(\vec x)$ and $b(\vec x)$ are regular functions
of $\vec x$. For neutral homopolymers in a good solvent, $v_0(\vec
x)$ is usually taken as a short range repulsive function (excluded
volume term). For
polyelectrolytes, $v_0(\vec x)$ is basically the Coulombic
interaction.  As for the random  ``charges'' $\{\xi_i\}$,  we will
consider them as independent random variables. Popular choices are the
binary or Gaussian forms.\par
More generally, one may consider that a monomer is
characterized by $z$ independent random ``charges''
$\xi_i^{\alpha}, (\alpha=1,2,..z)$ , linked to its electrostatic charge,
its hydrophilicity, its helix forming tendency,..... Of course, the
character of these ``charges'' can be chosen to be more complicated
(vectorial, tensorial,...) but we shall stick to the simple scalar
case. A very general model of heteropolymers may be then defined
through a two body interaction: 
\be
\label{free4}
v_{ij}(\vec r_i-\vec r_j)=\sum_{\alpha=1}^z c_{\alpha}
v_{ij}^{\alpha}(\vec r_i-\vec r_j) 
\ee
where $v_{ij}^{\alpha}(\vec r_i-\vec r_j)$ is defined through equation
(\ref {free3})
\be
\label{free5}
v_{ij}^{\alpha}(\vec r_i-\vec r_j)=v_0^{\alpha}(\vec r_i-\vec r_j)+\beta a^{\alpha}(\vec r_i-\vec
r_j)(\xi_i^{\alpha}+\xi_j^{\alpha})+\beta b^{\alpha}(\vec r_i-\vec
r_j)\xi_i^{\alpha}\xi_j^{\alpha} 
\ee
and $c_{\alpha}$ is a real number. Of course, this general model is
not very convenient to investigate, although it is clearly in the line
of the Hopfield model of spin glasses. For instance \cite{Ga_Ga_O}, it is possible to
show, that (i) if $b^{\alpha}$ is a zero range $\delta$ function (ii) if
$a^{\alpha}=0$ (iii) if one takes the limit $z\to
\infty$,  then the disordered contribution in the 
two body interaction  yields a random bond model that we consider
below. To illustrate some physical 
points, we now present a few simple disordered models. 

\subsection{Some basic models of disordered polymers}
\label{sec:Models}
\subsubsection{The random hydrophilic-hydrophobic chain}
\label{sec:HH}
In this model, one emphasizes the heterogeneity of the interactions
between the amino acids and the water, i.e. the solvent. 
In the context of protein folding, it is widely believed that the
interactions with the hydrophobic residues (Trp, Ile, Phe, ...)
are the main driving forces for the collapse of proteins: hydrophobic
residues in water can be thought of as being in a bad solvent, and
thus have an attractive effective interaction. The monomers
are described by a single ``charge'', $\xi_i$, and the monomer-solvent
interactions  are described by the Hamiltonian 
\begin{equation}
\label{diso1}
{\cal H}_{ms} = - \sum_{i=1}^{N} \sum_{\alpha=1}^{{\cal N}}  a ( {\vec
r_i}-{\vec R_{\alpha}} ) \xi_i
\end{equation}
where $a$ denotes a short-ranged monomer-solvent molecule
interaction, $N$ and $\cal N$ denote respectively the number of
monomers and of solvent molecules, and ${\vec r_i}$ and ${\vec
R_{\alpha}}$, their respective positions. Following appendix A, and
assuming that the system is incompressible, we have
\be
\label{diso2}
{\cal H}_{ms} = + \sum_{i=1}^{N} \sum_{j=1}^{N} a ( {\vec
r_i}-{\vec r_{j}})\xi_i - A \sum_{i=1}^{N} \xi_i
\end{equation}
where $A= \sum_{\vec r} a(\vec r)$.
The second term in (\ref {diso2}) is a constant, equal to $ N
A\overline{\xi}$, and will therefore be omitted henceforth.
We will write this hydrophilic-hydrophobic Hamiltonian in a
more symmetric way 
\be
\label{diso3}
\beta {\cal H}(\{v_{ij}\}) = {1 \over 2}\sum_{i=1}^{N} \sum_{j=1}^{N}  \left(v_0( {\vec  
r_i}-{\vec r_{j}})+ \beta a ( {\vec 
r_i}-{\vec r_{j}})(\xi_i+\xi_j) \right)
\end{equation}
which indeed looks like equation (\ref{free3}) with $b=0$. Up to this
point, we have not specified the probability distribution of the
disorder variables $\{\xi_i\}$. Striving for simplicity, we will write
\be
\label{diso4}
P(\{\xi_i\})=\prod_{i=1}^{N} h(\xi_i)
\end{equation}
namely we assume that the $\{\xi_i\}$ are independent random variables
drawn with the same probability law $h$. This problem will be studied
in section \ref{sec:hydro}.
\subsubsection{The random-bond chain}
\label{RBC}
This model play a central role in the mean-field theory of
heteropolymers, since its freezing phase transition is described by the
Random Energy Model. The two body interaction is given by
\be
\label{rb1}
v_{ij}(\vec r_i-\vec r_j)=v_0(\vec r_i-\vec r_j)+\beta w_{ij}(\vec r_i-\vec
r_j)
\ee
where the disorder contributions $w_{ij}$ are independent variables
(they are defined for $i<j$). This means that two couples $(i,j)$ and $(i',j')$
of the same chemical nature, but at different positions of the primary
sequence, are characterized by independent values of the
interactions $w_{ij}(\vec r_i-\vec r_j)$ and $w_{i'j'}(\vec r_{i'}-
\vec r_{j'})$ This model therefore assumes a very strong influence of
the environment on the heteropolymer. As mentioned above, it can also
be seen as a particular limiting case, where the number $z$ of
independent ``charges''
characterizing a monomer becomes very large.  Again looking for
simplicity, we will assume the probability distribution to be given by 
\be
\label{rb2}
P(\{w_{ij}\})=\prod_{(ij)=1}^{N(N-1)/ 2} h(w_{ij})
\end{equation}
where the condition $(i<j)$ is tacitly assumed.
This case will be studied in section \ref{sec:ranbon}. 
\subsubsection{The ``random sequence'' chain}
\label{sec:RS}
In marked contrast to the previous case, the random sequence chain
assumes that the monomers carry a single ``charge'' $\xi_i$ and that the
environment plays no role at all. The two-body interaction reads:
\be
\label{rs1}     
v_{ij}(\vec r_i-\vec r_j) = v_0(\vec r_i-\vec r_j) + \beta b(\vec
r_{i}- \vec r_j)\xi_i  \xi_j 
\ee
where $v_0$ again refers to excluded volume effects, and the second term
describes the random interaction. This term has very different
interpretations (and physical behavior) for $b>0$ and $b<0$. The
former allows one to think of equation (\ref{rs1}) as describing a
randomly (electrostatically) charged chain, or polyampholyte. The
latter describes random copolymers, e.g. AB copolymers , where
monomers of type A and B have a tendency to phase separate. If $b$
is constant, and notwithstanding the (crucial) chain constraint, the
spin analogs of these models are the antiferromagnetic $(b>0)$ or
ferromagnetic $(b<0)$ Mattis models. This is why the copolymer case
may illustrate how the low temperature phase may be linked with (or coded
in) the primary sequence ``charges'' $\{\xi_i\}$. These models are
considered in section \ref{sec:ranseq}, with the same assumption as in equation
(\ref{diso4}):
\be
\label{diso5}
P(\{\xi_i\})=\prod_{i=1}^{N} h(\xi_i)
\end{equation}
that is of the independence of the random variables {$\{\xi_i\}$}.
\par
We have presented very simplified models of disordered
self-interacting polymers, which may have some relevance to the
physics of protein folding. It is clear that many other polymer
problems are relevant to the physics of biopolymers. To name but a few, let us 
mention \par
(i) charged polymers subject to an electric field, in relation
to electrophoresis \cite{Zim_Lev} \par
(ii)  polymers at interfaces, in relation to membranes or adsorption
\cite{Ga_O,Sr_Ch_Sh}\par
(iii) polymers in random media \cite{Ed_Mu}. 

We now turn to a more detailed study of three ``basic''
heteropolymer models. 
\subsection{The random hydrophilic-hydrophobic chain}
\label{sec:hydro}
The Hamiltonian for this case has been derived above, and reads
\bea
\label{hamhyd}
\beta {\cal H}(\xi_i) =&& {1 \over 2} \sum_{i \ne j} \left [
v_0 + \beta ( \xi_i+\xi_j)\right]
 \ \delta(\vec r_i - \vec r_j) \nonumber\\
&&+ { 1 \over
6 } \sum_{i \ne j \ne k} w_0
\ \delta(\vec r_i- \vec r_j)\ \delta(\vec r_j - \vec r_k) \nonumber\\
&&+ { 1 \over 24 } \sum_{i \ne j \ne k \ne l} y_0
\ \delta(\vec r_i- \vec r_j)\ \delta(\vec r_j - \vec r_k)
\ \delta(\vec r_k - \vec r_l)
\eea
In equation (\ref{hamhyd}), we have assumed all interactions to be short
ranged, and we have also include three and four body terms for reasons
that will soon become clear \cite{Ga_O_Le}. 
\par
Following section \ref{sec:diso}, the partition function reads:
\be
\label{partit1}
Z(\{\xi_i\}) = \int \prod_{i}  d{\vec r_i} \prod_{i} 
g({\vec r_i},{\vec r_{i+1}}) \ \exp ( - \beta {\cal H}(\{\xi_i\}) )
\ee
leading to a disorder averaged free energy:
\be
\label{free11}
\overline{F} = - T\ \int \prod \ d\xi_i \ h(\xi_i) \log Z(\{\xi_i\})
\ee
Our calculations will be performed with the particular choice:
\be
\label{proba1}
h(\xi_i)={1 \over \sqrt{2 \pi \xi^2}}
\exp \left( - {(\xi_i-\xi_0)^2 \over 2 \xi^2 }\right)
\ee
With our conventions, a positive $\xi_0$ corresponds to a majority
of hydrophilic links.\par
Rewriting  (\ref{hamhyd}) and (\ref{partit1}) in a continuous form, we have:
\be
\label{partit2}
Z(\{\xi(s)\}) = \int {\cal D} {\vec r(s)} \exp \left( - {d \over 2 a^2 }\int_0^N ds
\left( {d {\vec r(s)} \over ds} \right)^2 - \beta{\cal H}(\{\xi(s)\}) \right)
\ee
with
\bea
\label{hamhyd1}
\beta {\cal H}(\{\xi(s)\})
= {1 \over 2}\int_0^N ds \int_0^N ds' \left(v_0 + \beta
(\xi(s)+\xi(s'))
\right)
\delta({\vec r(s)}-{\vec
r(s')})\nonumber\\ 
+ {w_0 \over 6} \int_0^N ds \int_0^N ds'\int_0^N ds'' \ 
\delta ({\vec r(s)}-{\vec
r(s')}) \ \delta ({\vec r(s')}-{\vec r(s'')})\nonumber\\
+ { y_0 \over 24}\int_0^N ds \int_0^N ds'\int_0^N ds'' \int_0^N ds'''
\ \delta({\vec r(s)}-{\vec r(s')})\ \delta({\vec r(s')}-{\vec
r(s'')})
\ \delta({\vec r(s'')}-{\vec r(s''')})
\eea
In the following, we will assume that the origin of the chain is fixed
at $\vec 0$ and that the extremity is free.
\par
We first consider the replica route, namely we consider typical
properties of a typical chain of the dilute ``soup'' of the previous section. 
Introducing replica indices $a,b,..$ we may perform the quenched
average in (\ref{free11}) and get 
\bea
\label{zav}
\overline{ Z^n} = \int \prod_{a=1}^n {\cal D} {\vec r_a(s)} \exp \left( - {d \over 2 a^2}
\int_0^N ds \sum_a
\left( {d {\vec r_a(s)} \over ds} \right)^2 - A \right)\nonumber\\
\times\exp \left( {\beta^2 \xi^2 \over 2}  \int_0^N ds \int_0^N ds'\int_0^N ds'' \ 
\sum_{a,b} \delta ({\vec r_a(s)}-{\vec
r_a(s')}) \ \delta ({\vec r_b(s)}-{\vec r_b(s'')}) \right)
\eea
with
\bea
\label{zav2}
A =&&{1 \over 2}(v_0 +2 \beta \xi_0 ) \int_0^N ds \int_0^N ds' \ 
\sum_a \delta({\vec r_a(s)}-{\vec r_a(s')})\nonumber\\  
&&+ {w_0 \over 6} \int_0^N ds \int_0^N ds'\int_0^N ds'' \ 
\sum_a \delta ({\vec r_a(s)}-{\vec
r_a(s')})
 \ \delta ({\vec r_a(s')}-{\vec r_a(s'')})\nonumber\\
&&+ {y_0 \over 24}  
\int_0^N ds \int_0^N ds'\int_0^N ds'' \int_0^N ds'''\nonumber\\
&&\sum_a\ \delta({\vec r_a(s)}-{\vec r_a(s')})\ \delta({\vec r_a(s')}-{\vec
r_a(s'')})
\ \delta({\vec r_a(s'')}-{\vec r_a(s''')}) 
\eea
\par
At this stage, there are two possibilities. The first one, due
to Edwards and Muthukumar \cite{Ed_Mu}, is a variational approach to the replicated
Hamiltonian of equations 
(\ref{zav},\ref{zav2}) through the trial Hamiltonian 
\be
\label{mutu}
\beta {\cal H}_0= {1\over 2}\int_0^N ds \int_0^N ds' \
\sum_{a,b} \left(\vec r_{a}(s)
g_{a,b}^{-1}(s-s')\vec r_{b}(s')\right)
\ee
This method is an extension of Des Cloizeaux calculation \cite{descl} for a
swollen polymer. The variational parameter(s) is the $n \times n$
matrix $g_{a,b}(s-s')$. 
This approach has the advantage that it does not rely on the ``ground
state dominance'' approximation (see appendix B and below), but it is
technically rather heavy. Beside the original work \cite{Ed_Mu}, which
delt with a polymer chain in a random medium, the only use of this
method, in the context of self-interacting random chains, is, as far
as we know, the hydrophilic-hydrophobic chain \cite{Mo_Ku_Da}. This
study deals with the three-dimensional case, in 
the framework of a one step replica symmetry breaking scheme for the
Parisi-like kernel $g(s,x), \  x\in [0,1]$. We have chosen to present
first a different approach, more mean-field in character, and then to
compare our results with this variational method.\par

The alternative route we have followed uses order parameters, namely
we introduce  the overlap $q_{ab}(\vec r, \vec r')$ , with $a < b$,
and the density $\rho_a (\vec r)$ by:
\bea
\label{param}
q_{ab}(\vec r, \vec r') &=& \int_0^N ds \ \delta ({\vec r_a(s)}- \vec r)
\ \delta ({\vec r_b(s)}-{\vec r'}) \nonumber\\
\rho_a(\vec r)&=& \int_0^N ds \ \delta ({\vec r_a(s)}- \vec r) 
\eea
we may write equations (\ref{zav},\ref{zav2}) as:
\bea
\label{zav3}
\overline{Z^n} = \int {\cal D} q_{ab}(\vec r, \vec r')
{\cal D} {\hat q}_{ab}(\vec r, \vec r') {\cal D} \rho_a(\vec r)
{\cal D} \phi_a(\vec r) 
\ \exp \left( G(q_{ab}, {\hat q}_{ab}, \rho_a, \phi_a) + \log 
\zeta({\hat q}_{ab}, \phi_a) \right)
\eea
where ${\hat q}_{ab}(\vec r, \vec r')$ and $\phi_a (\vec r)$ are the Lagrange
multipliers associated with (\ref{param}), and:
\bea
\label{zav4}
G(q_{ab}, {\hat q}_{ab}, \rho_a, \phi_a) = \int d^dr \sum_a \left( i \rho_a(\vec r)
\phi_a(\vec r) - (v_0 + 2 \beta \xi_0)\ { \rho_a ^2 (\vec r) \over 2}
- {w'_0 \over 6} \rho_a^3 (\vec r) - {y_0 \over 24}
\rho_a^4 (\vec r) \right) \nonumber\\
+ \int d^dr \int d^dr' \sum_{a < b} \left( i q_{ab}(\vec r, \vec r')
{\hat q}_{ab}(\vec r, \vec r') + \beta^2 \xi^2 \ q_{ab}(\vec r, \vec r')
\rho_a(\vec r) \rho_b(\vec r') \right)
\eea
and
\bea
\label{zeta}
\zeta ({\hat q}_{ab}, \phi_a) =&&
\int \prod_a {\cal D}\vec  r_a(s) \ \exp \left(-{d \over
2 a^2} \int^ N_0  ds\ \sum_a {\dot {\vec r_a}(s)}^2 -
i \int_0^N ds\ \sum_a \phi_a ( \vec r_a (s))\right)\nonumber\\
&&\times\exp \left(-i \int_0^N ds \sum_{a<b}\ {\hat q}_{ab}(\vec r_a(s),
\vec r_b (s))\right)  
\eea
In equation (\ref{zav4}), we have defined:
\be
\label{prima}
w'_0 = w_0 - 3 \beta^2 \ \xi^2
\ee
According to appendix B, we can rewrite:
\be
\label{quant1}
\zeta ({\hat q}_{ab}, \phi_a) = \int \prod_a d^d r_a 
<\vec r_1 \ldots \vec r_n | e^{-N H_n({\hat q}_{ab}, \phi_a)}
 | \vec 0 \ldots \vec 0>
\ee
where the $H_n$ is a ``quantum-like'' $n \to 0$ Hamiltonian, given by:
\be
\label{quant2}
H_n= -{ a^2 \over 2 d } \sum_a {\vec \nabla_a}^2 \ + \ \sum_a i \phi_a(\vec
r_a) + \sum_{a < b} i{\hat q}_{ab}(\vec r_a, \vec r_b)
\ee

So far our treatment has been rigorous.
Anticipating some kind of (hydrophobically-driven) collapse, we assume
that we can use ground-state dominance to evaluate
(\ref{quant1},\ref{quant2}), and write, omitting some non-extensive
prefactors (see appendix B):
\bea
\label{quant3}
\zeta ({\hat q}_{ab}, \phi_a) &&\simeq \ e^{-N E_0({\hat q}_{ab},
\phi_a)} \nonumber\\
&&\simeq \exp\left(-N\min_{\{\Psi(\vec r)\}}\left\{ <\Psi|H_n|\Psi> -
E_0 \left(<\Psi|\Psi>-1 \right)\right\} \right)
\eea
where $E_0$ is the ground state energy of $H_n$. 
At this point, the problem is still untractable, and we make the
extra approximation of saddle-point method (SPM). The extremization
with respect to $q_{ab}$ reads:
\be
\label{min1}
i{\hat q}_{ab}(\vec r,\vec r') = - \beta^2 \ \xi^2 \rho_a (\vec
r)\ \rho_b (\vec r')
\ee
This equation shows that replica symmetry is not broken
(at least at the saddle point level), since $\hat
q_{ab}$ is a product of two single-replica quantities $\rho_a$.
To get more analytic information, we follow appendix B, and use the
Rayleigh-Ritz variational principle. Due to 
the absence of replica symmetry breaking (RSB), we further restrict
the variational wave-function space to Hartree-like 
replica-symmetric wave-functions, and write:
\be
\label{var1}
\Psi(\vec r_1,\ldots,\vec r_n) = \prod_{a=1}^n \varphi(\vec r_a)
\ee
Because of replica symmetry, we can omit replica indices and 
take easily the $n \to 0$ limit.
The variational free energy now reads:
\bea
\label{var2}
-\beta \overline{F}(q, {\hat q}, \rho, \phi, \varphi) =\int d^dr \left( i \rho(\vec r)
\phi(\vec r) - (v_0 + 2 \beta \xi_0)\ { \rho^2 (\vec r) \over 2}
- {w'_0 \over 6} \rho^3 (\vec r) - {y_0 \over 24}
\rho^4 (\vec r) \right) \nonumber\\
-{1\over 2} \int d^dr \int d^dr' \left( i q(\vec r, \vec r')
{\hat q}(\vec r, \vec r') + \beta^2 \xi^2 \ q(\vec r, \vec r')
\rho(\vec r) \rho(\vec r') \right) \nonumber\\
- N \left\{\int d^d r
\ \varphi (\vec r) \left(-{a^2 \over 2 d } {\vec \nabla}^2 +  i \ \phi(\vec
r) \ \right) \varphi(\vec r)
 - {i \over 2} \int d^dr \int d^dr' 
\hat q(\vec r, \vec r') \varphi^2 (\vec r) \varphi^2 (\vec r') \right\}\nonumber\\
+N E_0 \left( \int d^dr \varphi^2(\vec r) -1 \right)
\eea
The SPM equations read:
\bea
\label{var3}
\rho(\vec r)&=&N\varphi^2(\vec r)\nonumber\\
q(\vec r,\vec r')&=&N\varphi^2(\vec r)  \varphi^2(\vec r') \nonumber\\
i\phi(\vec r)&=&(v_0 + 2\beta \xi_0) \rho(\vec r) + {w'_0 \over
2} \rho^2(\vec r) + { y_0 \over 6} \rho^3 ( \vec r) + \beta^2
\xi^2 \int d^dr' q(\vec r,\vec r') \rho(\vec r') \nonumber\\
i \hat q (\vec r,\vec r')&=&- \beta^2 \xi^2 \rho(\vec r) \rho(\vec
r') 
\eea
We still have to minimize with respect to the normalized wave-function
$\varphi(\vec r)$. This leads to a very complicated non-linear
Schr\"odinger equation, and we shall restrict ourselves to a
one-parameter family of Gaussian wave-functions of the form:
\be
\label{phi1}
\varphi(\vec r) =  \left( {1\over 2 \pi R^2} \right)^{d/4}
\exp(- { {\vec r}^2 \over 4 R^2})
\ee
where $R$ is the only variational parameter.
\par
Using equations (\ref{var1}) and (\ref{var2}),
the variational free energy per monomer ($ \overline{f} =
\overline{F}/N$) becomes:
\bea
\beta \overline{f} =&&{a^2\over2d} \int d^d r \left( {\vec \nabla}
\varphi (\vec 
r)\right)^2 + {(v_0+2\ \beta\ \xi_0)\over 2}\ N \ \int d^d r \varphi^4(\vec
r)\nonumber\\
&& +N^2 \left \{ {w'_0 \over 6} \int d^dr \varphi^6 (\vec r) +
{\beta^2 \xi^2 \over 2} \left(\int d^d r \varphi^4 (\vec r) \right)^2
\right \} + {y_0 \over 24} N^3 \int d^dr \varphi^8(\vec r)
\eea
Using some Gaussian integrals, and the value of $w'_0$
given in (\ref{prima}), the free energy reads:
\bea
\label{freevar}
\beta \overline{f}&=&{a^2\over8R^2}+ {1\over (2 \sqrt{\pi})^d}{(v_0+2\
\beta\ \xi_0)\over 2}{N\over R^d}\nonumber\\
& &+ \left( {1\over (2 \pi \sqrt{3})^d}
{w_0\over 6}- {1\over (2 \pi)^d}{\beta^2 \xi^2 \over
2}(3^{-d/2}-2^{-d}) \right) \  ({N\over R^d})^2 \nonumber\\
& &+ \left({1\over (32 \pi^3)}\right)^{d/2}
\ {y_0 \over 24}\  \left({N\over R^d}\right)^3
\eea
At low temperatures, one has to study  the sign of the third
term of (\ref{freevar}). The repulsive four body term is necessary to yield
a stable theory at low temperatures, when the coefficient of 
$\left( {N \over R^d } \right)^2$ changes sign, due to disorder
fluctuations. A detailed study \cite{Ga_O_Le} yields the following results:\par
(i) {\bf $\xi_0 > 0$ : the hydrophilic case.}
\par
In our approximation, we find that there is a first order transition
towards a collapsed phase ($R \sim N^{1/d}$) induced by a negative
three-body term (and stabilized by a positive four-body term). This
transition is neither an ordinary $\theta$ point (since the two-body
term is positive), nor a freezing point (the replica-symmetry is not
broken). Such an effect has been previously found in random
polyelectrolytes \cite{Qia_Kho}.
Note that since our approach is variational, the true free
energy of the system is lower than the variational one, and
we thus expect the real transition to occur at an even higher
temperature. Since the transition is first order, we expect
metastability and retardation 
effects to be important near the transition; we also expect that (due
to the latent heat), there will be a reduction of entropy in the low
temperature phase, compared to an ordinary second order $\theta$
point. Furthermore, the three body induced
collapse has some non trivial geometry built-in \cite{Brazo,Al_Ta}.
\par 
(ii) {\bf $\xi_0 < 0$ : the hydrophobic case.}
\par
In this case, defining the two temperatures:
\bea
\label{tempe}
T_0=& 2 |\xi_0| /v_0\nonumber\\
T_1=& \xi {\sqrt 3 \over \sqrt{w_0(1 - 3^{d/2}/2^d)}} 
\eea
yields two possible scenarii:\par
(iia) $T_1 > T_0$ : the collapse transition is again driven
by the disorder fluctuations of the three-body interactions. The
resulting first-order transition is very similar to the hydrophilic
case (i).\par
(iib) $T_0 > T_1$ : the collapse transition is now driven by
the strong two-body $\xi_0$ term. The resulting phase transition
is very similar to an ordinary $\theta$ point, and is therefore
second-order. At low temperature, the collapsed phase undergoes
another (first order) phase transition due to the negative three body term.

We therefore find that the random hydrophilic-hydrophobic chain has
two compact phases, separated by a first order phase transition.
One may also note that it has two swollen ``phases'', one without
metastable states 
(above the $\theta$ regime), and one with metastable states (above
the discontinuous collapse  regime). Note that, in a similar vein,
one can also find an extra collapsed regime (between the two
collapsed phases) displaying metastability effects.  These results
were obtained by a saddle point approach supplemented by a ground
state dominance approximation. The validity of these approximations is
discussed in appendix B, and we only mention here that this
approach is not really appropriate for the description of metastable
states. Lastly, it is of interest to note that annealed random chains
have a very similar phase diagram.\par  
At this point, it is interesting to compare our
results, which have essentially a mean field dignity, with the replica
variational approach of \cite{Mo_Ku_Da} in $d= 3$. This work
finds a phase diagram in broad agreement with ours, except that the
first order transitions become continuous `` one step '' freezing
transitions. One therefore gets in this approach the two swollen
phases mentioned above.
The fact that a replica symmetric mean field theory turns,
(for finite dimensions) into a broken replica symmetry theory
occurs also
in the random field Ising model \cite{Me_Mo}. Interestingly
enough, the random term in equation (\ref{hamhyd}) can be rewritten in
a ``random density'' way since 
\be
\label{rand}
\sum_{i,j} \xi_i \ \delta(\vec r_i - \vec r_j)= \sum_{i} \xi_i
\rho(\vec r_i)
\ee 
which, in some sense suggests that this random chain can also be
viewed as an Imry-Ma system. The Imry-Ma domain arguments depend on
non-extensive (free) energies and are not easily interpreted
 in the framework of replica theory. It would be much nicer
to work with a fixed $\{\xi_i\}$ distribution: in this respect, one
may think of a disorder dependent variational method, and/or a
domain size analysis. 
\par
 Finally, we close this section by some protein related comments. One may say
that ``good'' $\{\xi_i\}$ sequences  (i.e. easily foldable)
should not be 
trapped in a metastable state, and should yield stable geometrical shapes.
This means that a good sequence should enter neither of the two
swollen phases, since one has potentially trapping metastable states,
and the other leads to a $\theta$ collapsed phase which has no
definite shape (extensive conformational entropy). In this picture, a
good sequence should be, in some appropriate phase diagram, on the
dividing edge between the two swollen 
phases: the folding transition would then be a 
multicritical point, which seems reasonable from a physical
standpoint. This remark is probably related to some dynamical criteria
\cite{Ca_Thi,Shakdyn,Kli_Thi} that may characterize good folders in various
heteropolymer models.   \par 
\subsection{The random bond chain.} \par
\label{sec:ranbon}
There are various ways to treat this difficult case \cite{Gar_Orl1,Sh_Gu}, and we will
present our point of view along two lines. We will first consider the
very high dimensional approach, where the chain constraint is
irrelevant. In this case, one may show directly that a collapsed phase
undergoes a Random Energy Model (REM) freezing transition. One may
also follow the same path as for the
hydrophilic-hydrophobic chain, namely to use ground state dominance
plus some saddle point approximation. The main difference here is that
one is faced with a broken replica symmetry saddle point so that the
variational wave function is not replica symmetric.
\subsubsection{Very high dimension approach.}
\label{highdi}
Let us consider the Hamiltonian:
\be
\label{hamram}
\beta {\cal H}(\{w_{ij}\})={1 \over 2} \sum_{i\ne j}(v_0+\beta w_{ij}) \delta(\vec
r_i-\vec r_j)+ {1 \over 6} \sum_{i\ne j\ne k} w_0 \delta(\vec r_i-\vec
r_j)\delta(\vec r_i-\vec r_k)
\ee
where the couplings $\{w_{ij} \}$ are random independent couplings,
and
$v_0$ represents the overall effect of the solvent, as well as
the direct non random pair interactions. For the sake of simplicity,
we use a Gaussian probability distribution for the couplings:
\be
\label{probar}
h(w_{ij}) = {1 \over \sqrt{2 \pi w^2}} \exp{\left(- { w_{ij}^2 \over 2
w^2}\right)}
\ee
The partition function is:
\be
\label{partira}
Z(\{w_{ij} \}) = \int {\cal D}{\vec r}_i \prod_{i} g(\vec r_i,\vec r_{i+1})
e^{-\beta {\cal H}(\{w_{ij}\})}
\ee
where the function $g(\vec r_i,\vec r_{i+1})$ again enforces the chain
constraint. 
\par
Replicating and averaging (\ref{partira}) yields:
\bea
\label{lulu}
\overline{Z^n}&=&\int \prod_a
{\cal D}{\vec r}_i^a \prod_{i,a} g(\vec r_i^a,\vec
r_{i+1}^a ) \ \exp{\left(-\tilde v_0 \sum_{i<j} \sum_a \delta(\vec r_i^a - \vec
r_j^a )+ {\beta^2 w^2 \over 2} \sum_{a \ne b} \sum_{i < j}
\delta(\vec r_i^a -\vec r_j^a )
\delta(\vec r_i^b -\vec r_j^b )\right)}\nonumber\\
& &\times  \exp{\left(- {w_0 \over 6} \sum_{i \ne j \ne k} \sum_a
\ \delta(\vec r^a_i- \vec r^a_j)\ \delta(\vec
r^a_j - \vec r^a_k) \right)}
\eea
with $\tilde v_0 = v_0 - \beta^{2} {w^2 \over 2}$ .
\par
The above replicated Hamiltonian has three characteristics:\par
(i) the chain constraint term.\par
(ii) a possible $\theta$ point if $\tilde v_0 < 0$.\par
(iii) a possible freezing transition due to the $a \ne b$ term of
(\ref{lulu}).
\par
Since we wish to emphasize here the freezing transition, we will assume that
$\tilde v_0$ is indeed negative, so that the system is in the
collapsed phase. To get an easily tractable model, we further consider a simple
but unrealistic geometry, namely a collapsed chain on a fully
connected lattice. On such a lattice, by definition, each point is
a neighbor to all the other points of the lattice. Examples are
provided by a triangle ( 3 points, two dimensions), a tetrahedron (
four points, three dimensions),... so that for a large number $\Omega$
of points, it is a high dimensional ($\Omega -1$) polyhedron. On such
a lattice, the chain constraint i) is automatically satisfied, since a
site has $\Omega - 1$ neighbors, so that the chain constraint is a $1
/ \Omega $ effect. 
\par
Let us show that in the collapsed phase, the model is equivalent to a
Random Energy 
Model (REM). For that purpose, we will
show that the energies of this system are random independent Gaussian
variables. First, we note that in the
collapsed phase, the monomer density $\rho$ is finite and constant in
space; only a number $ N/\rho $ of sites are occupied. 
This implies that the only conformation dependent term of (\ref{hamram}) is
the random two-body term $\sum_{i<j} w_{ij} \delta(\vec r_i-\vec
r_j)$.
Since this term is a linear combination of Gaussian variables, it is
also a Gaussian variable, and thus, its distribution is entirely
characterized by its correlation functions. Therefore,
instead of
computing the joint probability $P(E_1,E_2)$ for two copies of the
chain, we will calculate directly the
correlation $\overline{E_1 E_2}$ between the energies $E_1$ and $E_2$
of two conformations $\{ \vec r_i^{(1)} \}$ and $\{ \vec r_i^{(2)} \}$
of the chain in the collapsed phase. We have:
\bea
\label{kiki2}
\overline{E_1^2} =& {w^2 \over 2} \sum_{i,j} \delta ( \vec r_i^{(1)}
-\vec r_j^{(1)})\nonumber\\
&= {w^2 \over 2} \sum_{\vec r} \rho^2_1(\vec r)\nonumber\\
&= {w^2 \over 2} N \rho 
\eea
where $\rho_1(\vec r) = \sum_i \delta(\vec r - \vec r_i^{(1)})$.
Therefore, $\overline{E_1^2}$ is independent of the (collapsed) 
conformation. Similarly, we have:
\bea
\label{kiki3}
\overline{E_1 E_2} =& {w^2 \over 2} \sum_{i,j} \delta ( \vec r_i^{(1)}
-\vec r_j^{(1)}) \delta ( \vec r_i^{(2)}
-\vec r_j^{(2)})\nonumber\\
&= {w^2 \over 2} \sum_{\vec r, \vec r'} q^2_{12}(\vec r,\vec r')
\eea
where $q_{12}(\vec r,\vec r') = \sum_i \delta(\vec r - \vec r^{(1)}_i)
\ \delta(\vec r' - \vec r^{(2)}_i) $. We have:
\be
\label{kiki4}
\sum_{\vec r,\vec r'} q_{12}(\vec r,\vec r') = N
\ee
and since all occupied points are equivalent (constant density), we
have:
\be
\label{kiki5}
q_{12}(\vec r,\vec r') = { \rho^2 \over N}
\ee
which in turn implies:
\be
\label{kiki6}
\overline{E_1 E_2} = {w^2 \over 2} \rho^2
\ee
from which we get the joint probability:
\be
\label{kiki7}
\lim_{N \to \infty} P(E_1, E_2) \sim \exp{\left( - { E_1^2 + E_2^2 \over  N
\rho w^2} \right)}
\ee
leading to a REM behaviour. This behaviour, in the context of protein
folding, was first put forward, on phenomenological grounds, by
Bryngelson and Wolynes \cite{Bry_Woly}.\par
It is of course possible to obtain these results using replicas. In
this case, one shows that the chain model 
as we have stated it, is equivalent to an infinite-range
Potts-glass, with $N / \rho \to \infty $ states. This model
can then be solved by a one-step replica symmetry breaking scheme.
The physics of the REM is discussed elsewhere in this
book: the chain undergoes a freezing transition at a temperature
$T_c$. Above $T_c$, the system has a finite entropy, whereas below
$T_c$, it vanishes. The system is then frozen in a small number of
dominant states, determined by subtle non extensive effects. This is
why this model is so appealing for protein folding.
\par
\subsubsection{High dimension approach.}
To go beyond the mean field REM freezing behaviour, one has to take
into account the chain constraint $ g(\vec r_i^a,\vec r_{i+1}^a )$ in
(\ref{lulu}). We have
\bea
\label{lulu1}
\overline{Z^n}&=&\int \prod_a
{\cal D}{\vec r}_i^a \prod_{i,a} g(\vec r_i^a,\vec
r_{i+1}^a ) \ \exp{\left(-\tilde v_0 \sum_{i<j} \sum_a \delta(\vec r_i^a - \vec
r_j^a ) + {\beta^2 w^2 \over 2} \sum_{a \ne b} \sum_{i < j}
\delta(\vec r_i^a -\vec r_j^a )
\delta(\vec r_i^b -\vec r_j^b )\right)}\nonumber\\
& &\times \exp{\left(- {w_0 \over 6} \sum_{i \ne j \ne k} \sum_a
\ \delta(\vec r_i^a - \vec r_j^a )\ \delta(\vec
r_j^a - \vec r_k^a ) \right)}
\eea
that we now rewrite as
\bea
\label{lulu2}
\overline{ Z^n}&=&
\int \prod_{a=1}^n {\cal D} {\vec r_a(s)} \exp \left( - {d \over 2 a^2
}
\int_0^N ds \sum_a
\left( {d {\vec r_a(s)} \over ds} \right)^2 - A \right)\nonumber\\
& &\times \exp \left( {\beta^2 w^2 \over 2}
 \int_0^N ds \int_0^N ds' \ 
\sum_{a< b} \delta ({\vec r_a(s)}-{\vec
r_a(s')}) \ \delta ({\vec r_b(s)}-{\vec r_b(s')}) \right)
\eea
with
\bea
\label{lulu3}
A =&&{1 \over 2}(\tilde v_0 ) \int_0^N ds \int_0^N ds' \ 
\sum_a \delta({\vec r_a(s)}-{\vec r_a(s')})  \nonumber\\
&&+ {w_0 \over 6} \int_0^N ds \int_0^N ds'\int_0^N ds'' \ 
\sum_a \delta ({\vec r_a(s)}-{\vec
r_a(s')}) \ \delta ({\vec r_a(s')}-{\vec r_a(s'')})
\eea
\par
Defining, as in equation (\ref{param}), the parameters $q_{ab}(\vec
r, \vec r')$ , with $a < b$, and $\rho_a (\vec r)$, and introducing
the associated Lagrange multipliers ${\hat q}_{ab}(\vec r, \vec r')$
and $\phi_a (\vec r)$, one gets:
\bea
\label{lulu5}
\overline{Z^n} = \int {\cal D} q_{ab}(\vec r, \vec r')
{\cal D} {\hat q}_{ab}(\vec r, \vec r') {\cal D} \rho_a(\vec r)
{\cal D} \phi_a(\vec r) 
\ \exp \left( G(q_{ab}, {\hat q}_{ab}, \rho_a, \phi_a) + \log 
\zeta({\hat q}_{ab}, \phi_a) \right)
\eea
with
\bea
\label{lulu6}
G(q_{ab}, {\hat q}_{ab}, \rho_a, \phi_a) =&& \int d^dr \sum_a \left( i \rho_a(\vec r)
\phi_a(\vec r) - (\tilde v_0 )\ { \rho_a ^2 (\vec r) \over 2}
- {w_0 \over 6} \rho_a^3 (\vec r) \right) \nonumber\\
&& + \int d^dr \int d^dr' \sum_{a < b} \left( i q_{ab}(\vec r, \vec r')
{\hat q}_{ab}(\vec r, \vec r') + {\beta^2 w^2 \over 2} \ q_{ab}^2(\vec r, \vec r')
\right)
\eea
and
\bea
\label{lulu7}
\zeta ({\hat q}_{ab}, \phi_a) =&&\int{\cal D}\vec  r_a(s)
 \exp \left(-{d \over 2a^2} \int^ N_0  ds\ {\dot {\vec r_a}}^2 \right)
\nonumber\\
&&\times \ \exp\left(-i \int_0^N ds\ \sum_a \phi_a (\vec r_a (s))
-i \int_0^N ds \ \sum_{a<b}{\hat q}_{ab}(\vec r_a(s), \vec r_b (s)) \right)
\eea
To go further, one follows the same approximations as in section
 (\ref{sec:hydro}).\par 
(i)  one assumes ground state dominance in the ``quantum''
Hamiltonian
associated with equation (\ref{lulu7}).\par
(ii) the free energy is calculated by the SPM.
\par
In view of the high dimension results, point (i) is natural, since one
expects first a $\theta$ collapse transition, followed at low
temperature by a freezing transition driven by the off-diagonal (in
replica space) terms. The procedure exactly parallels the one of the 
randomly hydrophilic-hydrophobic chain of the preceding section,
except for the variational wave function that enters the
Rayleigh-Ritz principle. The replica symmetric form extracted from equation
(\ref{min1}) is not valid anymore: the variational wave function
should present replica symmetry breaking. A simple form was proposed
by Shakhnovich and Gutin \cite{Sh_Gu}, and reads:
\be
\label{psivar}
\Psi(\vec r_1,\ldots,\vec r_n) = {{({\rm det} \ K)^{d/4}} \over {(2 \pi)^{nd /4}}}
\exp \left( {- {1 \over4} \sum_{a,b}\vec r_a K_{ab} \vec r_b} \right)
\ee
where $K$ is a $n \times  n$ Parisi-like hierarchical matrix (see
\cite{Me_Pa_Vi}), and $d$ is the
dimension of space. The variational free energy is extremized with
respect to $K$. The result, for large enough $d$, is a step function
 form for $K(x), \ (x\in [0,1])$, corresponding to a REM type of replica symmetry
breaking. We briefly recall the physical meaning of a
$x$-dependent length scale (see equation (\ref{psivar})). The overlap
parameter of equation (\ref{param}) can be understood by considering
two real chains $\vec r_{1}(s)$ and $\vec r_{2}(s)$, with the same
disorder configuration $\{v_{ij}\}$ coupled through an infinitesimal term
of the form
\be
\label{coup}
H_{12}= \varepsilon \int_0^N ds \ \delta(\vec r_1(s)-\vec r_2(s))
\ee
It can be shown, as in spin glasses, that the Parisi order parameter
$q(x)=\lim _{n\to 0} q_{ab}(\vec r, \vec r)$ is identical
to the average $q_{12}=<\delta(\vec r_1(s)-\vec r_2(s))>$ taken over the two
(real) chains Hamiltonian. Small $x$ corresponds to the average overlap
over large distance (of the order of the global radius of gyration),
whereas $x\sim 1$ corresponds to an average overlap over a microscopic
distance (of the order of a bond length). In this picture, the $x$
dependence stems from the existence of many different local minima in
the two chain system. These minima can be probed with an infinitesimal
field $\varepsilon$.\par
It is unclear to us whether this analysis applies for $d=3$, but a
variational wave function of the form (\ref{psivar}) has been widely
used in other disordered condensed matter situations \cite{Ba_Bou_Me}:
vortex lattice with impurities in superconductors, interface in a
random potential,  amorphous solidification of vulcanized
macromolecules,...\par 
The previous approaches to the random bond chain study a freezing
transition in the collapsed r\'egime. Using the Edwards-Muthukumar
approach \cite{Ed_Mu}, it should be possible to study also the direct
transition swollen phase $\to$ frozen phase. This transition is seen
in low dimensional simulations, yielding a phase diagram in large
agreement with the one of the preceding section. For essentially the
same reasons as in section \ref{sec:hydro}, one may define ``good folders''
as heteropolymers undergoing the folding transition at a multicritical
point.
\par
\subsection{The ``random sequence'' chain}
\label{sec:ranseq}
We now turn to the last case, namely the ``random sequence'' chain \cite{Gar_Orl2.Sh_Gu2}
As mentioned in section \ref{sec:diso}, this model
describes two very different physical realities, namely randomly
charged (globally neutral) polymers (also called polyampholytes) and random
AB copolymers.  The random variables $\{\xi_{i}\}$ of equation (\ref{rs1})
will be assumed to be independent. We first present a general strategy
for the two cases along the lines of section \ref{sec:hydro}:
\subsubsection{The ``standard'' approach}
\label{standard}
Consider a ``random sequence'' chain described by a two body term:
interaction:
\be
\label{rano}
v_{ij}(\vec r_i-\vec r_j) = v_0(\vec r_i-\vec r_j) + \varepsilon \beta  b(\vec
r_{i}- \vec r_j)\xi_i  \xi_j 
\ee
where the probability distribution $h(\xi)$ is given by equation
(\ref{proba1}), with $\xi_0=0$, and $\varepsilon=1$ 
(resp. $\varepsilon=-1$) applies to the 
polyampholyte (resp. copolymer) case. Following the same route as
above, we get:
\bea
\label{poly1}
\overline{Z^n} = \int {\cal D} q_{ab}(\vec r, \vec r')
{\cal D} {\hat q}_{ab}(\vec r, \vec r') {\cal D} \rho_a(\vec r)
{\cal D} \phi_a(\vec r) {\cal D} \Psi_{a}(\vec r)  
\ \exp \left( G(q_{ab}, {\hat q}_{ab}, \rho_a, \phi_a, \Psi_a) + \log 
\zeta({\hat q}_{ab}, \phi_a) \right)
\eea
where ${\hat q}_{ab}(\vec r, \vec r')$ and $\phi_a (\vec r)$ are again
the Lagrange multipliers associated with (\ref{param}), and:
\bea
\label{poly2}
G(q_{ab}, {\hat q}_{ab}, \rho_a, \phi_a, \Psi_a) = \int d^dr \sum_a \left( i \rho_a(\vec r)
\phi_a(\vec r) - v_0 \ { \rho_a ^2 (\vec r) \over 2}
- {w_0 \over 6} \rho_a^3 (\vec r) - {\xi ^{2}\over 2} \rho_a(\vec
r)\Psi^2_a(\vec r)\right) \nonumber\\
- {1\over {2\beta \varepsilon}}\int d^dr \int d^dr'\Psi_a(\vec
r)b^{-1}(\vec r-\vec r')\Psi_a(\vec r')\nonumber\\
+ \int d^d\int d^dr' \sum_{a < b} \left( i q_{ab}(\vec r, \vec r')
{\hat q}_{ab}(\vec r, \vec r') - \xi^2 \ \Psi_a(\vec r) q_{ab}(\vec r, \vec
r')\Psi_{b}(\vec r') \right)
\eea
and
\bea
\label{poly3}
\zeta ({\hat q}_{ab}, \phi_a) =
\int \prod_a {\cal D}\vec  r_a(s) \ \exp ((-{d \over
2 a^2} \int^ N_0  ds\ \sum_a {\dot {\vec r_a}}^2 -
i \int_0^N ds\ \sum_a \phi_a ( \vec r_a (s))\nonumber\\
\exp (-i \int_0^N ds \sum_{a<b}\ {\hat q}_{ab}(\vec r_a(s), \vec r_b (s))))
\eea

Clearly, one may run through the same analysis as was done before
(saddle point method plus ground state dominance approximation). It
is easy to see that the SPM yields a replica symmetric solution, for
the same reasons as in section \ref{sec:hydro}. For instance, the
saddle point equation for $\Psi_a(\vec r)$ reads 
\be
\label{poisso}
-{1\over {2\beta \varepsilon}}\int d^dr'b^{-1}(\vec r-\vec
r')\Psi_a(\vec r')- \xi^2 \left(\rho_a(\vec r)\Psi_a(\vec r) +\int d^dr'\sum
_{b\ne a}q_{ab}(\vec r, \vec
r')\Psi_{b}(\vec r')\right)=0
\ee
At this stage, two different cases have to be considered. For short
range forces, ((a) random AB copolymer chain, (b) polyampholyte chain
with salt), the 
saddle point equations yield either a macroscopic phase separation
(case (a)) or a macroscopic charge crystallization (case (b)). This
clearly shows that in this case, the SPM is a rather poor
approximation and does not treat properly the chain constraint. For
case (a) above, it can easily be shown that a macroscopic phase separation
would only occur in the irrealistic fully 
connected geometry of section \ref{highdi}.\par
\subsubsection{Beyond the saddle point method: short range interactions}
\label{beyond}
The solution to this
problem has been found by Leibler \cite{Lei} in the case of a {\it
melt} of (non-random) AB block  
copolymers: when one goes beyond the SPM, there appears a new length
scale $l^{*} \sim \sqrt{N}$, which is the spatial scale for phase
separation between A-rich and B-rich regions ($N$ is the length of one
copolymer chain). In other words, the order parameter Fourier 
component $\Psi_a(\vec k)$ has critical fluctuations, for all wave
vectors on a sphere of radius $ \vert\vec k^{*}\vert= {2\pi \over l^{*}}$.
Due to this continuous symmetry, it was shown by Brazovskii
\cite{Brazo} that, below $d=6$, thermodynamic fluctuations turn this
transition into a first order transition. Depending upon the content
of A's and B's in the chains, one expects various types of modulated
ordered phases (lamellar, hexagonal, body centered cubic,...) for the
order parameter $\Psi_a(\vec r)$. The random AB {\it melt} presents
the same type of modulated order, added to a possible replica symmetry
breaking phenomenon induced by the order parameter $ q_{ab}(\vec r,
\vec r')$ of equation (\ref{poly1}). A similar situation arises for
the short range polyampholyte melt. The discussion of the ``random
sequence'' melt is therefore quite complicated, due to the interplay
of a non zero wave vector ordering and of disorder.  Since the 
techniques and results are specifically linked to the notion of a
polymeric melt,  we refer the interested ready to the literature
\cite{Dobry,Shak,Anger}, where phase diagrams involving modulated
and/or frozen structures have 
been proposed. Lastly, the dynamics of this model may help to bridge 
the gap with real glasses  \cite{Ki_Thi}, since the peculiar features of 
Brazovskii's phase transition lead to a large number of symmetry
unrelated metastable phases above the critical temperature.\par
As a temporary conclusion to the disordered short range
interactions case, we wish to make some remarks which pertain to the
{\it single} chain problem.\par
(1) for a melt, the existence of the wave vector $\vec k^{*}$ can be
established  either through replica calculations or directly (see equation
(\ref{free1})): we see no 
reason why the associated length scale, in the single chain problem,
should not be distributed, as is frequently found in domain
arguments. \par
(2) for screened Coulomb interaction (polyampholyte with salt),
numerical evidence \cite{Imb,Kard,Grass} supports the existence of a
$\theta$ transition for 
a neutral chain in $d=3$. As far as we know, no evidence exists for a
modulated or frozen phase at very low temperatures. Finally we mention
that a one dimensional version of a directed chain shows a very non
trivial freezing transition \cite{der_hi}.\par 
\subsubsection{Beyond the saddle point method: long range interactions}
\label{long}
For long ranged	interactions (polyampholytes without salt), the saddle
point equation (\ref{poisso}) is indeed reminiscent of the
Poisson-Boltzmann equation. Again the chain constraint would be poorly
treated at this stage, and one would have to go beyond the SPM as for
the short range case. Moreover, it is not clear that the free energy of
the chain is self averaging in the presence of Coulomb forces. We will
therefore appeal to (more classical) electrostatic arguments
( see \cite{Hi_Jo,Gu_Sh,Do_Ru,Pa_Gro} and references therein).\par 
Consider a soup of neutral polyampholyte chains of $N$ monomers, each
of length $l$. For each chain, monomer $i$
is given a charge $\xi_i=\pm q_0$, with probability $h(+)=h(-)=1/2$ (of
course one has $b(\vec x)={1\over {\vert\vec x\vert}^{d-2}}$).\par
For reasons that will soon become clear, one has to distinguish two
different situations, both of experimental interest: one may either
consider\par 
(i) an ensemble of random chains where the neutrality
constraint holds separately for each chain or \par
(ii) an ensemble of random chains which are globally neutral, implying that
each chain has, up to a random sign, a typical charge $Q \simeq
q_0\sqrt {N}$.\par

In the former case, the low temperature phase is collapsed, since, as
shown by Higgs and Joanny \cite{Hi_Jo}, one then gains an
electrostatic condensation energy of order $-Nq_0^2/l^{d-2}$. The
transition is very similar to an ordinary
$\theta$ point with renormalized excluded volume. Furthermore there
are some numerical evidence \cite{Pa_Gro} that there may exist, at
still lower temperature, a freezing transition of the REM type.\par 
In the latter case, the excess charge $Q$ may lead to a swelling of
the chains. Its associated 
electrostatic energy $E_Q$ is of order $Q^{2}/R^{d-2}$, where $R$ is
the radius of gyration. \par
This suggests that above $d=4$, this electrostatic
contribution may be irrelevant since we then have $R \sim N^{1/2}$,
implying $E_Q \sim N^{(4-d)/2}$. One then expects a situation very
similar to case (i) above.  
Below $d=4$,  the two cases are different. The role of the Coulomb
interaction in $d=3$ is not yet settled, although there are some
evidence that the behaviour of the chain is controlled by a parameter
$(\alpha = Q/Q_R)$, where $Q_R$ is the Rayleigh charge well known in the
instability of a spherical charged droplet. For $\alpha$ small, the
polyampholyte chain is collapsed at low temperature, whereas it is
swollen for $\alpha$ large, possibly into an elongated necklace of
collapsed beads. Again, a freezing transition is still possible at
lower temperature.  
\par

\subsection{Conclusions}
We have presented some of the problems linked with the ``random
sequence'' melt, where modulated and/or frozen phases may appear, that
one may interpret in term of (more or less) finite range phase
separation or charge condensation. These phases show up beyond the
saddle point approximation. The long range character 
of Coulomb interactions may be a further difficulty. The discreteness
of the chain can also be the source of complications (commensurability
effects,...). 
The relevance of these results for a single chain (and may be for
protein folding) is unclear, to say the least. It is tempting to assume
that similar considerations apply once the chain has undergone a
collapse $\theta$ transition, but we believe that new theoretical methods 
must be found for the one chain problem. On the protein side, only
five (His, Lys, Arg, Asp, Glu) 
among twenty of the amino acids are charged and their position along
the chemical sequence are somehow correlated \cite{Pan_Gro_Ta}.
Furthermore, the rather small size of realistic proteins probably
make any freezing ``transition'' a non REM ``transition'' \cite{Pa4}.
\subsection{Heteropolymers and proteins}
\label{sec:hetero}
A general feature of the above heteropolymer models is the existence of several
compact phases, with widely different entropies (and other
characteristics as well). We have also pointed out that one may have
different coil phases, with widely different dynamical
behaviour. The theoretical methods we have presented can certainly be
criticized, and we recall some of their weak points:\par
(i) the results are obtained in the limit $N \to \infty$.\par
(ii) we have basically used the SPM supplemented by a ground state
dominance approximation. The need to go beyond the SPM is clear in the
``random sequence'' melt, but the single chain problem seems presently
out of reach. Furthermore, the ground state approximation is not
appropriate to describe the coil $\to$ globule transition.\par
(iii) we have made calculations over disorder averaged quantities,
instead of considering a fixed disorder configuration. For a single
chain, the difference can be important, and an approach along the
lines of appendix B should be interesting. Furthermore, we have
assumed, for simplicity, that the disorder variables where
uncorrelated, which is certainly not realistic. It is therefore of
interest to mention a recent extension \cite{Plot_W} of the REM to
take energy correlations into account.  
\par
As for proteins, there is a general agreement about their being
non-random, either from a hydrophobic point of view \cite{Ir_pet}, or from a
Coulombic point of view \cite{Pan_Gro_Ta}. This non randomness has been
also emphasized in a dynamical context
\cite{Ca_Thi,Shakdyn,Kli_Thi}. Nevertheless, the comparison 
between heteropolymers and proteins has proven a useful idea: one may
mention, among other features, the existence of a molten globule
\cite{Pti} as another distinct compact phase of proteins, the
successful interpretation of some (thermo)dynamical folding
experiments by REM or other glassy models (see
e.g. \cite{Frauen,You_Pow,Eaton,Angell}), or the 
tentative design of simplified folding potentials, of fast folding
sequences (for recent references see e.g. \cite{Deu_Ku,Seno})..... We
emphasize once more that, from the heteropolymer 
point of view, a ``good'' folder should follow, in some appropriate
phase diagram, a dividing edge from the coil state onto a folded
state, through a multicritical folding transition.  

\section{Dynamics of proteins}
In this section, we shall review some theories of protein
dynamics. Roughly speaking, one can distinguish
two main approaches to this problem. 

The first approach, of a microscopic
nature, is based on the dynamics of the polymer models discussed in
the previous sections. It thus encompasses the 
topological frustration induced by the
chain constraint, and the steric hindrance.
Considering the difficulties of 
the thermodynamics,
this approach is still at a very preliminary
stage, especially in the collapsed phase, where entanglements and
topological frustration effects are dominant.
To further emphasize the difficulty of the problem, let us mention that there
is, at present, no satisfactory theory for the dynamics of the
collapse of a homopolymer chain. As one can imagine, the problem is
orders of magnitude harder for heteropolymer chains.

The second approach, of a rather
phenomenological nature, relies on the strong similarities
between the protein folding problem and the random energy model (REM),
and reduces somehow to variants of the dynamics of the REM. This
approach concentrates on the roughness of the energy landscape due to
energetic frustration.

In a first section, following the microscopic route
we study the dynamics of collapse of a
homopolymer chain. This may be useful to describe the first stages of
the hydrophobic collapse of a protein, since it does not
crucially depend on the specific nature of the interactions.
We also briefly discuss 
various attempts towards a
description of heteropolymer dynamics along these lines.

Finally, we present some dynamical approaches to the REM, as an
oversimplified protein folding scheme.

\subsection{Dynamics of the collapse of a polymer chain}

In this section, we study the
dynamics of protein folding.
At the very early stages of protein folding, the driving force
is believed to be the hydrophobic force; it is thus reasonable, in a
first approximation,  to
neglect the amino acid sequence, and model the system as a homopolymer
in a bad solvent.

According to de Gennes' theory \cite{PGG1,PGG2} the collapse of a
flexible coil leads to the formation of ``pearls''  on a minimal scale
along the linear chain, which thickens and shortens
under diffusion of the monomers, then
forms new pearls at a larger scale,
until the final state of a compact globule
is reached; the longest timescale for the collapse,
neglecting knot formation, is
estimated as 

$$\tau_c \sim {\eta a^3 \over k_B  \theta } \left({\theta \over|\Delta T|
}\right)^4  N $$
where $\eta$ is the viscosity of the solvent, $\theta$ is the
temperature of the $\theta$ point
, $a$ is the monomer size and $\Delta T$ is the temperature
quench from the $\theta$ temperature. This time $\tau_c$ has a strong dependence on 
molecular
weight. For the case of proteins (see reference \cite{BILL}), $N=300$,
this yields a collapse time of $\tau_c \sim 1 $ s for a temperature quench of 
${\Delta T \over \theta} = 0.01$.

In a series of articles, Timoshenko  et al.
\cite{TIM1}
have developed an alternative theory based on 
a self-consistent method using Langevin equations
that can be analyzed numerically; kinetics laws for the collapse
of a homopolymer are obtained with or without 
hydrodynamic interactions, at early and later stages.
In a recent article, a generalization of this method
has been applied to the dynamics of a hydrophilic-hydrophobic
heteropolymer.

In the following, we shall present an analytical method to study
the kinetics of a homopolymer in a $\theta$ solvent when it is quenched
into bad solvent
conditions (collapse into a globule) \cite{SPhT}.

\subsubsection{Presentation of the method}

We consider a homopolymer chain in $\theta$ conditions - i.e a Gaussian coil-
consisting of N monomers, obeying  the Langevin dynamics as 
the chain is quenched 
into good or bad solvent conditions (equations (\ref{1}) and (\ref{2}) ).

Let's first neglect all hydrodynamic interactions.
To keep the notations as simple as possible, we 
will omit the  arrows on the vectors.
The
equations of motion for the system read:

\begin{eqnarray}
    && \zeta\frac{\partial{r}}{\partial{t}} = - 
        \frac{\partial{H}}{\partial{r}} + \eta(s,t) \label{1}\\
    && H=\frac{3k_B T}{2 a_{0}^{2}} \int_{0}^{N}
        \left(\frac{\partial{r}}{\partial{s}}\right)^{2}ds +V(r(s,t)) \label{2}
\end{eqnarray}
where $N$ is the total number of monomers, 
$r(s,t)$ is the position of monomer $s$ in the chain,
$a_{0}$ is  the monomer length and $\zeta={k_B T \over D}$
is the friction coefficient, $D$ 
is the diffusion constant
of a monomer in the solvent and $k_B T$ is the temperature. The
intra-molecular as well as intermolecular interactions of the
chain are contained in the potential $V(r(s,t)) $.

The thermal noise $\eta(s,t)$ is a Gaussian noise with zero
mean and correlation given by:
$$<\eta(s,t)\eta(s',t')>=2\zeta k_B T \delta(s-s') \ \delta(t-t')$$

The method consists in finding a virtual homopolymer chain
which obeys a simpler Langevin equation,
chosen so that its
radius of gyration best approaches the radius of
gyration of the real chain at each time $t$.

The virtual chain, defined by $r^{(v)}(s,t)$ satisfies the Langevin equation:

\begin{eqnarray}
    && \zeta\frac{\partial{r^{(v)}}}{\partial{t}} = -
        \frac{\partial{H_{v}}}{\partial{r^{(v)}}} + \eta(s,t) \label{3}\\
    && H_{v}=\frac{3 k_B T}{2 a^{2}(t)} \int_{0}^{N}
       \left (\frac{\partial{r^{(v)}}}{\partial{s}}\right)^{2}ds  \label{4}
\end{eqnarray}
with the same friction coefficient and noise as the original equation,
but with a simpler Hamiltonian $H_v$. Indeed this Hamiltonian
$H_v$ represents  a Gaussian chain, but with a time dependent  Kuhn
length $a(t)$.

Our method is a generalization of Edwards' 
uniform expansion model \cite{EDW} to dynamics. This method consists
in 
calculating the radius of gyration
of a polymer by using perturbation theory, and
adjusting the simplified Hamiltonian so that the first order
perturbation to the radius of gyration vanishes.
If $v$ denotes the excluded volume,
the method gives the Flory radius \cite{Flory} for large $N$
and agrees with the result of the first-order perturbation 
expansion for small $v$. Note that it would seem natural to use
the most general quadratic Hamiltonian rather than that of (\ref{4})
, but this was shown by des Cloizeaux
\cite{desCloizeaux} to yield the incorrect exponent $\nu = 2/d$.

Let's define
$$ \chi(s,t)=r(s,t)-r^{(v)}(s,t)$$
$$W= H - H_v$$
Assuming that (\ref{3}) is a good approximation
to (\ref{1}) , $\chi(s,t)$ and $W$ can be regarded as small,
and to first order in these quantities, the dynamical equations
become:

\begin{eqnarray}
    && \zeta\frac{\partial{r^{(v)}}}{\partial{t}} = 
        \frac{3 k_B T}{a^{2}(t)} 
            \frac{\partial^{2}{r^{(v)}}}{\partial{s^{2}}} +\eta(s,t) \label{5}\\
   && \zeta\frac{\partial{\chi}}{\partial{t}} = 
         \frac{3 k_B T}{a^{2}(t)} 
            \frac{\partial^{2}{\chi}}{\partial{s^{2}}}
+3 k_B T\left(\frac{1}{a_{0}^{2}}-
      \frac{1}{a^{2}(t)}\right)\frac{\partial^{2}{r^{(v)}}}{\partial{s^{2}}}
  +F(r^{(v)}(s,t)) \label{6}
\end{eqnarray}
where $F(r(s,t))=-\frac{\partial{V}}{\partial{r(s,t)}}$
is the driving force for the swelling or collapse of the chain.

More precisely, in the following,
for a chain in a bad solvent,
we will take attractive two-body interactions and repulsive
three-body interactions:

 \begin{eqnarray*}
&& V(r(s,t))=-V_{2}(r(s,t)) +V_{3}(r(s,t))\\
&& V(r(s,t))=-\frac{v}{2} k_B T\int_{0}^{N}ds\int_{0}^{N}ds'
        \delta(r(s,t)-r(s',t))\\
 &&  +\frac{w}{6} k_B T\int_{0}^{N}ds\int_{0}^{N}ds'\int_{0}^{N}ds''
        \delta(r(s,t)-r(s',t)) \delta(r(s',t)-r(s'',t)) , 
 \end{eqnarray*} 
where $v>0$  and $w>0$.

In this approximation, the radius of gyration of the chain
becomes:
\begin{eqnarray}
  R_{g}&&=\frac{1}{N}\int_{0}^{N}<r^{2}(s,t)>ds  \\
    &&\simeq \frac{1}{N}\int_{0}^{N}<((r^{(v)})^{2}(s,t) + 2 r^{(v)}(s,t) \chi(s,t))>ds  
\label{7}
\end{eqnarray}
The brackets denote the thermal average ( that is an average over the
Gaussian noise $\eta(s,t)$).
Our approximation consists in choosing the parameter $a(t)$ in such a
way that the first order in (\ref{7}) vanishes:

\begin{equation}
   \int_{0}^{N}<r^{(v)}(s,t)\chi(s,t)>=0 \label{8}
\end{equation}
or in Fourier coordinates:
\begin{equation}
    \sum_{n\neq 0}<\tilde{r}^{(v)}_{n}(t)
                    \tilde{\chi}_{n}^{*}(t)>=0  \label{9}
\end{equation}
where the Fourier transform is given by:
 \begin{eqnarray*}  
&& \left\{\begin{array}{ll}  
                 \tilde{r}_{n}(t)=\frac{1}{N}
         \int_{0}^{N}e^{i\omega_{n}s}r(s,t)ds\\  
                 r(s,t)=\sum_{n\neq 0}
      e^{-i\omega_{n}s} \tilde{r}_{n}(t) 
          \end{array} \right. \\ 
 \end{eqnarray*} 
and similarly for $ \chi(s,t)$.

We have used periodic boundary conditions, so that 
$\omega_{n}=\frac{2\pi n}{N}$.
In addition, to get rid of the center of mass diffusion, we constrain
the center of mass of the system to remain at fixed position,
$\tilde{r}_{0}(t)=
 \tilde{r}^{(v)}_{0}(t)
=\tilde{\chi}_{0}(t)=0$.

Equations (\ref{5}) and (\ref{6}) can easily be solved in Fourier
space.
We assume that at time $t=0$, the chains are in a $\theta$ solvent, so
that the initial condition $\{r(s,0)\}$ obeys Gaussian statistics. We
choose the initial virtual chain to coincide with the real one, so
that $r^{(v)}(s,0) = r(s,0)$ for any $s$. Denoting by
$\overline{\cdots}$ the average over the initial conditions,
the correlation function of $r(s,0)$ (in Fourier space) is taken as:

\begin{eqnarray}
\overline{\tilde{r}_{n}(0)} &&= 0 \\
\overline{\tilde{r}_{n}(0)\tilde{r}_{m}^*(0)}
  && =\frac{Na_{0}^{2}}{4\pi^{2}n^{2}}\delta_{mn} 
\end{eqnarray}

In Fourier space, the thermal noise is characterized by:
\begin{eqnarray}
 <\tilde{\eta}_n (t)> && =0\\
 <\tilde{\eta}_n (t)\tilde{\eta}_m^{*}(t')>&& =
\frac{2\zeta k_B T}{N} 
        \delta_{nm} \delta(t-t'). 
\end{eqnarray}

Replacing 
$\tilde{r}^{(v)}_{n}(t)$ and
$ \tilde{\chi}_{n}^*(t)$ by their expression
in (\ref{9}), and taking thermal and initial condition 
averages, we obtain an implicit equation for $a(t)$.


This equation can be solved analytically in both
limits
$ t << \tau_R $ (short time limit) and $ t >> \tau_R $
(long time limit)
where $\tau_R = {N^2 a_0^2 \over 4 \pi^2 D}$
is the Rouse time.

In order to take into account the hydrodynamic interactions with the solvent,
one has to modify the Langevin equations in the following way \cite{DOI};
equations (\ref{1}) and (\ref{3}) have to be replaced by:

\begin{eqnarray}
    && \frac{\partial{r(s,t)}}{\partial{t}} = 
          \int_{0}^{N}ds' {\bf O}(r(s,t)-r(s',t))   
        \left[-\frac{\partial{H}}{\partial{r(s',t)}} + \eta(s',t)\right]\\
    && \frac{\partial{r^{(v)}(s,t)}}{\partial{t}} = 
          \int_{0}^{N}ds' {\bf O} (r^{(v)}(s,t)-r^{(v)}(s',t))   
        \left[-\frac{\partial{H_{v}}}{\partial{r^{(v)}(s',t)}} + \eta(s',t)\right]
\end{eqnarray}
where ${\bf O}(r)$ is the Oseen tensor:

$${O}_{\alpha\beta}(r)=\frac{1}{8\pi\eta r} 
       \left(\delta_{\alpha\beta} 
     + \frac{r_{\alpha}r_{\beta}}{r^{2}}\right),$$
and $\eta$ is the viscosity of the solvent.

\subsubsection{Results}
In the absence of hydrodynamics interactions the results are the following:
at short times $t << \tau_R$, the radius
of gyration decreases as a power law:

\begin{equation}
R_{g}^{2}(t)=Na_{0}^{2}(1-\sqrt{\frac{t}{\tau_{c}}})
\end{equation}
with a characteristic time $\tau_c$ defined in (\ref{tauc})
and for large times $t>> \tau_c$, the radius of gyration relaxes to
that of a compact globule, resulting from the competition
between the two-body and three-body terms, according to

\begin{equation}
R_{g}(t)\sim (\frac{w}{v})^{\frac{1}{d}}
            N^{\frac{1}{d}}(1+e^{-\frac{t}{\tau_{2}}}),
\end{equation}
where
$$\tau_{2}\sim {1\over 4 \pi^2 D} (\frac{w}{v})^{\frac{2}{d}} N^{1+\frac{2}{d}}. 
$$

Note that for dimensions larger than $2$, 
the relaxation time is much shorter than the Rouse time.
For example in $d=3$,  $\tau_{2}\sim
N^{\frac{5}{3}}$ compared to $N^2$

The first stage of  collapse can be
characterized by a time scale
$\tau_{c}$
given by:

\begin{equation}
\tau_{c}^{\frac{1}{2}}=
\frac{8(4\pi)^{\frac{d}{2}}a_{0}^{d+1}}
     {\sqrt{6D \pi}vI_{d}N^{\frac{2-d}{2}}}
\label{tauc}
\end{equation}
where
$$
I_{d}=\int_{0}^{1}du
        \int_{0}^{1}du'
       \frac{1}{[|u-u'|(1-|u-u'|)]^{\frac{d}{2}}}  \  \  \ {\rm with}
\ \ |u-u'| > \Lambda$$
and $\Lambda=1/N$ is a short distance cut-off.

For $d<2$, the integral converges for small 
$\Lambda$, and $I_d$ is independent of $N$.
On the other hand, 
for $d\geq2$, the integral is infra-red divergent and 
thus there is an
explicit dependence on the cut-off.
It is easily seen that this $N$ dependence exactly cancels
out the $N$ dependence in (\ref{tauc}) so that the final
characteristic time $\tau_c$ is finite (independent of $N$).
In particular, for $d=3$, 
we find
$$\tau_{c}\sim \frac{64\pi^{2}}{3 D}
    \left(\frac{a_{0}^{3}}{v}\right)^{2}
      \left({a_{0} \over 5.22}\right)^{2}$$.

The order of magnitude of this short time collapse can be
calculated for a typical protein in water. The diffusion constant
of a single amino-acid in water is typically
$D\sim 10^{-5}{\rm cm}^{2}/s$. As pointed out in reference \cite{BILL}, a
monomer unit in a protein consists of approximately 5 amino acids (due
to chain stiffness). Thus a typical value for the Kuhn length is $a_0
= 7 \AA$.
Consequently, the number of monomer units
is 30 for a chain of 150 aminoacids.
We find a microscopic characteristic time
$\tau_{c}\sim 10^{-8}s$ , the Rouse time being
$\tau_R \sim 10^{-6}$ s. Note that the relaxation
time $\tau_2 \sim 10^{-6} s$ is of the order of magnitude of the Rouse time.
The microscopic time $\tau_c$ is several orders of magnitude lower
than other estimates in the literature ( see references  \cite{BILL,thirum}). 

If one includes the hydrodynamic interactions,
the results are modified as follows:
at short times $t<<\tau_Z$ where
$\tau_Z=\pi \left(\frac{1}{ 2\pi}\right)^{\frac{3}{2}}
\frac{\eta a_0^3}{k_B T}  N^{\frac{3}{2}}$
is the Zimm time.  
The radius
of gyration decreases again like a power law, but with a
smaller exponent
than in the absence of hydrodynamic backflows:
\begin{equation}
R_{g}^{2}(t)=Na_{0}^{2}(1-\left(\frac{t}{\tau_{c,h}}\right)^{\frac{1}{3}})
\end{equation}
where the characteristic time $\tau_{c,h}$ is defined by
$$\tau_{c,h}^{\frac{1}{3}}=
9 \pi^{\frac{d}{2}} (2\pi)^{\frac{3}{2}}
\left(\frac{\pi\eta}{2 k_B T}\right)^{\frac{1}{3}}
\frac{a_0^{d+1}}{v N^{\frac{3-d}{2}}J_d}.$$

The integral $J_d$ is given by:
$$
\label{Jd}
J_d=\int_0^1du \int_0^1du'\sum_{p\geq1}
\frac{1-cos(2\pi p (u-u'))}{p^{\frac{5}{2}}}
\frac{1}{[|u-u'|(1-|u-u'|)]^{1+\frac{d}{2}}}
\  \  \ {\rm with}
\ \ |u-u'| > \Lambda$$
and $\Lambda=1/N$ is a short distance cut-off.

The singularity of (\ref{Jd}) can be analyzed, by noting that
for $x\rightarrow 0^+$,
$$\frac{\sqrt{2}}{\pi}\Gamma\left(\frac{1}{2}\right)
\sum_{p\geq1}
\frac{1-cos(2\pi p x)}{p^{\frac{5}{2}}}\sim (2\pi x)^{\frac{3}{2}}.$$

Therefore, for $d<3$, the integral converges for small 
$\Lambda$, and $J_d$ is independent of $N$.
On the other hand, 
for $d\geq3$, the integral is infra-red divergent and 
thus there is an
explicit dependence on the cut-off,
more precisely,
$ J_d\sim N^{\frac{d-3}{2}}.$

Note that again, $\tau_{c,h}$ does not depend on 
the number of monomers $N$.

Numerically, we find  $\tau_{c,h}\sim 10^{-8}s$
for a protein in water,
and the Zimm time is $\tau_{Z}\sim 10^{-6}s$.

For larger times , the radius of gyration relaxes to
that of a compact globule according to

\begin{equation}
R_{g}(t)\sim (\frac{w}{v})^{\frac{1}{d}}
            N^{\frac{1}{d}}(1+e^{-\frac{t}{\tau_{2,h}}}),
\end{equation}
where
$$\tau_{2,h}\sim 
\pi \left(\frac{1}{ 2\pi}\right)^{1+\frac{1}{d}}
\frac{\eta a_0^3}{k_B T}  N^{\frac{3}{d}}.  
$$

\subsubsection{Summary}

We have presented here a model where explicit attractive hydrophobic
forces have been introduced for the collapse of the protein chain.
We have assumed that the protein specificity is
not important at the beginning of the collapse,
and that the dynamics is the same as for a homopolymer chain.
Our main result is that the chain collapses locally on a time scale
$\tau_c \sim 10^{-8}$ s,
with or without the hydrodynamic interactions with the solvent.
This time is smaller than other theoretical
estimates and doesn't depend on the number of
amino-acids.
This shows
that in the early stage, the collapse is indeed a very local phenomenon,
where nearby amino-acids aggregate into small domains.

We emphasize again that in the large time regime, this calculation 
(as well as all other microscopic calculations) is
not fully reliable, since it widely underestimates the
entropy. The Flory theory predicts
a zero entropy for the collapsed phase, whereas it is known to be
extensive \cite{descl}. As a consequence, this method as well as
others, cannot account correctly for the conformational changes
between the various collapsed configurations. As a result, the
relaxation times are orders of magnitude smaller than those predicted
by more phenomenological theories \cite{PGG2}.

Recently, there have been some promising attempts to extend the microscopic
approach to the dynamics of  random heteropolymers
\cite{TIM6,Thirum_Ash_Bhat,Roan_Shakh}. From a technical point
of view, these approaches avoid the use of replicas (but not the
quenched averages). Since these calculations are quite involved, we
refer the reader to the original papers.

Experimentation in this field is quite difficult
since one has to work in a very dilute regime
in order to avoid aggregation of chains
and truely see the hydrophobic collapse.
The most promising experiments are those by
Chan et al. \cite{BILL}  who study sub-millisecond protein folding
by ultra-rapid mixing;
based on optical techniques, these experiments 
can monitor folding
up to the microsecond time-scale.

\subsection{Phenomenological approach}
We have seen in the previous section that under certain conditions,
random heteropolymer models can be viewed as REM. As described in
reference \cite{Bry_Woly}, the native state of a protein
corresponds to the lowest energy state of the system, and all states
with non native contacts have a higher energy. Due to the diversity 
of the amino acids, certain non native contacts can
be favorable, whereas others might be unfavorable. It is thus natural
to assume that the low lying states of a protein can be described by a
REM. As we have discussed in section \ref{sec:hetero}, this assumption of a REM
is reasonable in the collapsed phase of the protein, but certainly not
in its coil phase.

Although the
analogy of heteropolymer models with REMs is purely thermodynamical
and has no deep dynamical significance, it is tempting to pursue this
analogy for the dynamics. One is thus led to model the dynamics of
folding of a protein by the dynamics of a REM
(\cite{Dominicis_Laine_Orland,Koper_Hilhorst,Shakhnovitch_Gutin,Wolynes_Onuchic} 
and references therein). This approach neglects
the geometrical frustration that the system encounters when evolving
from one conformation to another, but mimics the energetic frustration.

We follow the method presented in \cite{Dominicis_Laine_Orland,Koper_Hilhorst}.
We use a master equation to describe the transitions between the
various states of the system, and design the transition rates in such
a way that the system eventually relaxes to equilibrium. It should be
noted that in some systems, there are cases where the
system never relaxes to equilibrium, although the transition rates
satisfy detailed balance. This phenomenon, related to aging, is
treated in the review by Bouchaud et al\cite{Bouchaud}. We
believe that this phenomenon is not present in proteins, due to their
small sizes.

The master equation we use is:
\be
\label{master}
{dP_{\alpha} \over dt} = \sum_{\gamma} \left( W_{\alpha \gamma}
P_{\gamma}(t) - W_{\gamma \alpha }
P_{\alpha}(t) \right)
\ee
where $P_{\alpha}(t)$ denotes the probability for state $\alpha$ with
energy $E_{\alpha}$ to be occupied at time $t$, and $ W_{\alpha
\gamma}$ is the transition probability from state $\gamma$ to state
$\alpha$ per unit time.

To guarantee that the equilibrium distribution of the system will be a
Boltzmann distribution, it is sufficient that the transition rates
satisfy the detailed balance relation:
\be
\label{detailed}
{ W_{\alpha \gamma} \over W_{\gamma \alpha }} = e^{-\beta (E_{\alpha}
-E_{\gamma})} 
\ee
We will discuss later in detail the choice of the transition rates. However,
one important point is that in a globular phase (for which
the REM has some significance), the $W_{\alpha \gamma }$ should not
connect any state to any other state. 

To simplify notations, we will consider a REM with $N$ degrees of
freedom and $2^N$ states, instead of the more general form $q^N$ of
section \ref{sec:ranbon}.
The probability for
the system to have an energy level of energy $E$ is given by:
\be
\label{probap}
P(E) = {1 \over \sqrt{ \pi N J^2}} e^{-{E^2 \over NJ^2}}
\ee
where $J$ is an appropriate energy scale.
As was discussed in section \ref{sec:ranbon}, such a model has a freezing
transition at temperature $T_c =  {J \over 2 \sqrt{\log 2}}$.

The typical energies scale like  $J \sqrt{N}$, and it is not
reasonable to assume a finite transition rate for a direct transition
from any state to any other.

We are thus led to assume that the system was prepared at not too high
temperature (i.e.at finite excitation energy from its ground state),
and that it will relax by making transitions between the lowest lying
states of the system. This is an important assumption, but we believe
that it is crucial in representing proteins by REMs.

If we restrict ourselves to the lowest lying states of the REM, it is
well known that the distribution of energies of the system is not
Gaussian anymore, but rather exponential \cite{Me_Pa_Vi}. 
Indeed, we are not sampling the
bulk of the probability distribution, but rather its low energy tail
(rare events). The calculation goes as follows. Let us denote by $M$
the number of states that we want to include in the dynamics. This
corresponds to the occupation of  the $M$ lowest energy states of the
REM. In other words,
one may remove all energies above a certain threshold energy $E^*$,
such that:
\be
\label{M}
M=2^N \int_{-\infty}^{E^*}  dE P(E)
\ee
The integral of $P(E)$ yields an error function:
\be
{1\over 2} \left(1 + \erf\left({E^* \over \sqrt{NJ^2}}\right)\right) = {M\over 2^N}
\ee
where the error function is defined by:
\bea
\erf(z)&=&{2\over \sqrt{\pi}} \int_0^z dt e^{-t^2} \\
&\sim& 1-{1 \over \sqrt{\pi} z} \ e^{-z^2}\ \  {\rm if}\ \  z \to +\infty
\eea
Since we take ${M \over 2^N} \to 0$, we obtain:
\be
\label{E*}
E^* = E_0(N) +{J
\log(2M\sqrt{\pi\log 2}) \over 2\sqrt{\log 2}} + O(1)
\ee 
where $E_0(N)$ is the ground state energy of the REM
\be
E_0(N)= -NJ\sqrt{\log 2} +{J \log N \over 4 \sqrt{\log 2}} 
\ee
The term involving the number
of states $M$ considered represents the low lying levels.
 The energies of these states, denoted by $\{E_\alpha\}$,
are independent random variables between
$E_0$ and $E^*$. Linearizing the Gaussian distribution around the
ground state, we find that these low lying states are
distributed according to the exponential law:
\bea
\label{calp}
{\cal P}(E_\alpha) &=& \beta_c e^{\beta_c (E_\alpha-E^*)} \ \  {\rm
if} \ \ E_\alpha  \le
E^* \nonumber \\ 
&=& 0  \ \  {\rm if} \ \ E_\alpha > E^*
\eea
where
\be
\beta_c = {1\over T_c} 
\ee
and 
\be
T_c = {J \over 2 \sqrt{\log 2}}
\ee
is the critical temperature of the REM. 

There remains the question of the sensitivity of the method with
respect to the choice of the number of states$M$. From (\ref{E*}),we obtain:
\be
\label{v}
v=M e^{-\beta_c (E^*-E_0)}={1\over 2 \sqrt{\pi\log 2}}
\ee
independent of $M$. So, we can safely take the limit where the
number of states $M$ goes to infinity, and adjust the threshold energy
$E^*$ so that equation (\ref{v}) is satisfied.

We are thus led to study the dynamics of a set of random independent
energy levels, distributed according to an exponential distribution
law (\ref{calp}).

To completely specify the dynamics of the system, we must define
the transition probabilities. The choice is not unique of course, but
we will be guided by some physical considerations. The main assumption
we will make is that the transition rate of the system between two states
is completely dominated
by the largest barrier that the system encounters between the two states.
As we mentioned
above,
in order to use
REM types of models, the system must be in a globular phase. We have
seen in the previous section that under certain conditions, there may
be a ``molten globule'' phase in the high temperature region, and a
folded or native state in the low temperature phase. Our goal is to
study the
dynamics in both phases.

The choice of transition rates is governed by the detailed balance
equation (\ref{detailed}). Following \cite{Koper_Hilhorst}, we take a
general form:
\be
\label{W}
W_{\alpha \gamma} = \Gamma_0 \ e^{-\beta \epsilon_{\alpha}} V_{\alpha} V_{\gamma}
\ee
where the $\epsilon_{\alpha}$ are the random non-extensive parts of the random
energies $E_{\alpha} = E_0 (N)+\epsilon_{\alpha}$ and the $V_{\alpha} >0 $
can be viewed as barriers (this point will be discussed in the
following). The constant $\Gamma_0$ is the inverse of the transition
time scale.

Without specifying the $V_{\alpha}$, the master equation
(\ref{master}) can be solved by taking a Laplace transform.
Defining the Laplace transform of $P_{\alpha}$ by:
\be
\tilde{P_{\alpha}}(z) = \int_0^{\infty} e^{-tz} P_{\alpha}(t)
\ee
the master equation can easily be solved. Indeed, defining:
\bea
\tilde{Q}(z) &=& \sum_{\alpha} V_{\alpha} \tilde{P}_{\alpha}  \\
\zeta &=&  \sum_{\alpha} e^{-\beta \epsilon_\alpha} V_{\alpha}
\eea
we find:
\be
\tilde{Q}(z) = {1\over 1 - \Gamma_0 \sum_{\alpha} {e^{-\beta
\epsilon_{\alpha}} V_{\alpha}^2 \over z+\Gamma_0 V_{\alpha} \zeta}} 
\sum_{\alpha} {V_\alpha P_{\alpha}(0) \over  z+\Gamma_0 V_{\alpha}
\zeta}
\ee
from which we get
\be
\label{Ptilde}
\tilde{P_{\alpha}}(z)= {P_{\alpha}(0) + \Gamma_0 e^{-\beta
\epsilon_{\alpha}} V_{\alpha} \tilde{Q}(z) \over  z+\Gamma_0
V_{\alpha} \zeta}
\ee
where $P_{\alpha}(0)$ is the initial probability of occupation of
state $\alpha$. 

The function $\tilde{P_{\alpha}}(z)$ is meromorphic, and thus $P_{\alpha}(t)$
can be obtained by the inverse Laplace
transform:
\be
\label{P(t)}
P_{\alpha}(t) = \int_{C-i\infty}^{C+i\infty} {dz\over 2 i \pi} e^{zt}
\tilde{P_{\alpha}}(z)
\ee
where $C$ is a real constant to the right of the largest pole of
$\tilde{P_{\alpha}}(z)$. This integral can be performed by using the
residue method:
\be
P_{\alpha}(t)=\sum_{{\rm all \ poles}} {\rm Res}(\tilde{P_{\alpha}},
z_p) e^{z_p t}
\ee
where ``Res'' denotes the residue and $z_p$ the value of the pole.

Various other time correlation functions can be calculated along
these lines.
We are thus led to study the pole and residue
structure of (\ref{Ptilde}). 

\begin{figure}
\centerline{
\psfig{figure=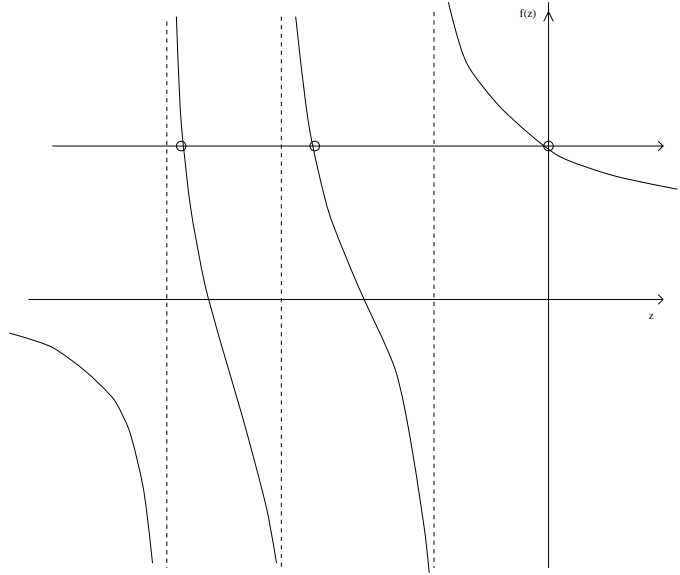,height=3.0in} }
\caption{Graphical solution of the pole equation
\label{F1}}
\end{figure}

Apart from the obvious pole $z_p= - \Gamma_0
V_{\alpha} \zeta$, the other poles are the solutions of the equation:
\be
\Gamma_0 \sum_{\alpha} {e^{-\beta
\epsilon_{\alpha}} V_{\alpha}^2 \over z+\Gamma_0 V_{\alpha} \zeta}= 1
\ee
This equation can be solved graphically (see Fig.~\ref{F1}).

We see that there is a pole at $z_p=0$, corresponding to the static
(infinite time ) limit. In addition, one sees that all poles are
negative, and close to the values $z_{\alpha} = -\Gamma_0 V_{\alpha}
\zeta$. To analyze further the explicit time dependence of
(\ref{P(t)}), it is necessary to specify the values of the
$V_{\alpha}$, and to perform the quenched average over the energy
levels distribution.

From the above analysis, we see that as $M\to \infty$, the poles will
concentrate between -$\Gamma_0 V_{max}
\zeta<0$ and $0$, and eventually give rise to 
a cut in the complex plane. The time dependence of $P_{\alpha}(t)$ is
controlled by the behaviour of the density of poles around
$z_p=0$.

In the following, we will assume that the system undergoes two phase
transitions: one at $T_{\theta}$ , similar to a theta point, from a swollen to a disordered
collapsed phase, and one at lower temperature $T_c$ , between the disordered
collapsed phase and the native phase. As mentioned above, the REM makes
sense only in the globular region, so that we restrict our
description to that region. Since the chain constraint is not present
in this approach, there is no induced geometry in the phase space. The
following choices of transition rates mimic in some sense a topology
in phase space.

\subsubsection{Model I: the high temperature region: $T_c < T < T_{\theta}$ }
In the high temperature phase, we expect the system to have a single
well-defined minimum, corresponding to a ``liquid'' condensed
phase. This may correspond to the 
so-called ``molten-globule'' phase
\cite{Pti}.

\begin{figure}
\centerline{
\psfig{figure=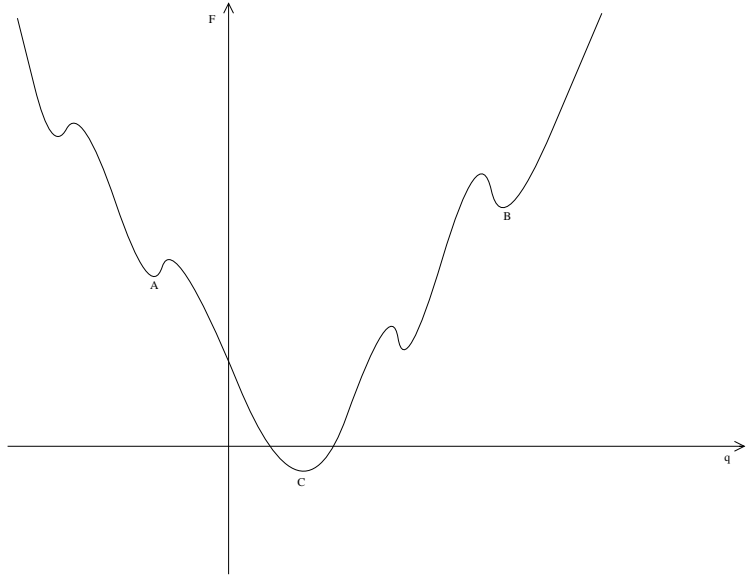,height=3.0in}}
\caption{Schematic phase space of model I
\label{F2}}
\end{figure}

In addition, there should exist many local minima due to the
existence of entanglements barriers.
The corresponding phase space is schematically represented in
Fig.~\ref{F2}. 

The point labeled C represents this liquid phase.
The typical path for a transition from B to A is
downhill from B to C then uphill from C to A. The
dominant barrier is thus from C to A, and it is natural to assume an
Arrhenius like law:
\bea
W_{A B} &=& W_0 \exp(-\beta(E_{A}-E_C)) \\
  &=&  W_0\exp(-\beta(\epsilon_{A}-\epsilon_C)) \\
\eea
The transition probability depends only on the final state $\alpha$ of the
system. This amounts to take $V_{\delta}=V $ for all states $\delta$ in
equation (\ref{W}) and to take $W_0 = \Gamma_0 V^2 e^{-\beta \epsilon_ C}$.
This model has been studied in  \cite{Dominicis_Laine_Orland}. Although
observables decay exponentially for a given instance of energy levels, 
for large times, the disorder averaged 
observables relax to their equilibrium
value like stretched exponentials:
\be
\overline{P_{\alpha}(t)} \simeq 
\overline{P_{\alpha}^{\rm eq}} + 
\overline{( P_{\alpha}(0)  -  P_{\alpha}^{\rm eq})}
\exp \left( -v_{\theta} \Gamma (1- {\beta_{\theta} \over \beta}) (W_0 t)^{\beta_{\theta} \over
\beta} \right)
\ee
where $\Gamma$ is the gamma function, $\beta_{\theta}$ is the inverse
theta temperature of the system, and $v_{\theta}$ is given by equation
(\ref{v}), with $\beta_{\theta}$ instead of $\beta_c$

In this liquid condensed phase of a globular
protein, dynamical processes are slowed down, due to the large
number of metastable states on the way to the ground state. 
The exponent of the stretched exponential is equal to 
$T/T_{\theta} \le 1$, and becomes equal to
1 at the theta temperature.
For temperatures
close to the theta temperature, it is close to 1  and
thus the dynamics looks close to a standard exponential relaxation.

Such a stretched exponential behaviour has been observed in
simulations of random heteropolymer models over several order of
magnitudes of the time \cite{Iori_Marinari_Parisi}.

The phase
space is reminiscent of what is defined as a ``funnel'' by
\cite{Wolynes_Onuchic}, and this suggests that the dynamics through a funnel
should be stretched exponential. Let us note also that the
phase space assumed here is similar to the one expected in each valley of the low
temperature phase (see
below).

\subsubsection{Model II: the low temperature region:  $T < T_c $}
\begin{figure}
\centerline{
\psfig{figure=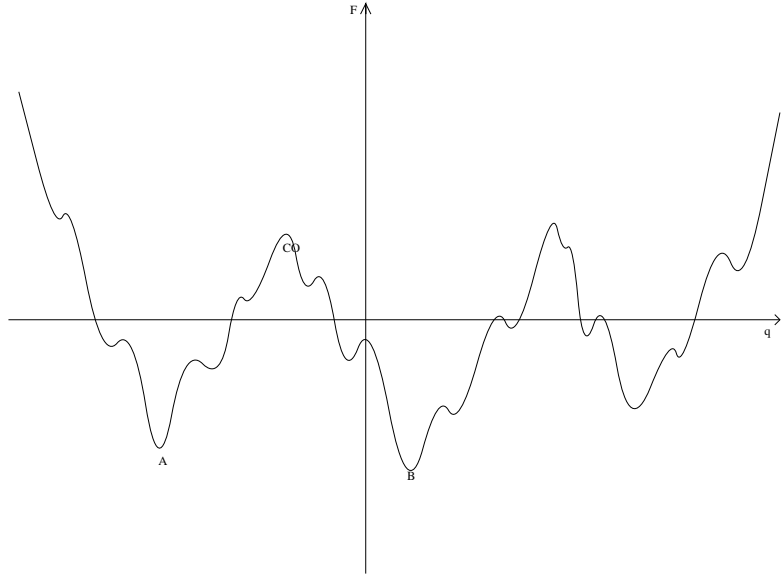,height=3.0in}}
\caption{Schematic phase space of model II
\label{F3}}
\end{figure}

In the low temperature phase, the system is expected to have a rugged
landscape, with many quasi-degenerate minima, corresponding to
incorrectly folded proteins. This corresponds to the phase space of
the native protein.
This phase space is
schematically represented in Fig.~\ref{F3}.

 As in the high temperature
case, the path from a low lying state B to A is through a barrier,
represented by C. Assuming that the height of the barrier between two
states is essentially constant, we take the following form for the
transition amplitudes:
\bea
W_{A B} &=& W_0 \exp(-\beta(E_C-E_{B})) \\
  &=&  W_0\exp(-\beta(\epsilon_C-\epsilon_{B}))
\eea
The transition probability depends only on the initial state $\gamma$
of the
system, and more precisely on the barrier from the initial state to
some typical point $C$ through which the system should pass. This
corresponds to the special case $V_{\delta}= V\ e^{\beta
\epsilon_{\delta}}$ in equation (\ref{W}), 
with $W_0 = \Gamma_0 V^2 e^{+\beta \epsilon_ C}$.

This model was studied by Koper and Hilhorst
\cite{Koper_Hilhorst}. The quenched average of (\ref{P(t)}) can be
performed, and the analysis of the long time behavior of the
probability yields algebraic decay at large times:
\be
\overline{P_{\alpha}(t) - P_{\alpha}^{\rm eq}}  \simeq  t^{-{T\over T_c}}
\ee

This slow relaxation
in the phase space of the native state, is due to the existence of
many quasi-degenerate local minima.

\subsubsection{A model for chaperon assisted protein folding}
It has been shown recently \cite{chaperon} that in living cells, 
there are specific
enzymes which catalyze the folding of proteins. 
These catalytic proteins have been named chaperones;
they were first discovered while
heating up some bacteria (hence their original name of
``heat-shock'' proteins). Each chaperon can help certain specific
proteins to fold. Although there is no consensus on the mechanisms
through which they act, their universal character (they can catalyze
the folding of many different proteins) suggests that they recognize
some major differences between a native protein and a misfolded one. As
we have seen in the previous sections, the collapse transition is
driven by the hydrophobic effect: the protein folds so as to maximize
the number of hydrophobic residues in its inner core, and the number
of hydrophilic or charged residues on its outer shell. Presumably, 
misfolded proteins will have a larger number of hydrophobic residues on
the outside, and a larger number of hydrophilic residues in the
inside. The presence of these hydrophobic patches on the outside of
the misfolded protein could
then be recognized by a chaperon.

Let us discuss a model that was originally proposed by Todd et al.
\cite{Todd}. The idea is that the chaperon protein binds to the
misfolded protein (by aggregating to their hydrophobic patches), and
transfers high amounts of energy to the misfolded protein (through ATP
hydrolysis). This ``energy kick'' pushes the protein out of its
misfolded minimum, and gives it a chance to relax back 
to the native
state. 
In addition, the chaperon does not bind to the native protein,
(since it has an optimal hydrophobic outer shell, with no apparent
hydrophobic patch).

This model has been studied in the master equation formalism defined
above \cite{Orland_Thirumalai}. The random energy states are the
misfolded states, 
and the
chaperon provides energy to induce transitions between these
states. In addition, the native state, which we denote $O$, is not
recognized by the chaperon. Thus, once the protein is in the native
state, it will stay there forever.

From the point of view of the transition rates $W_{\alpha \gamma}$, we
use the same model as in equation (\ref{W}), except that we impose
that the native state $O$ is not connected to any other state:
\be
W_{\alpha O} = 0
\ee
Since misfolded proteins can refold into the native state,
$W_{O \alpha}$ (which is the
transition rate from any state $\alpha$ to $O$) has no reason to
vanish. These assumptions seem to violate detailed balance
(\ref{detailed}). We can however ignore this fact by noting that
folding in presence of a chaperon is not really a thermal
equilibrium phenomenon, and that these assumptions amount
to assume that the free energy of the native state is much lower than
that of all misfolded states (large gap).

We may write two equations, one for the occupation of the native state
$O$ and one for any misfolded state $\alpha$:
\bea
\label{Pt}
{d \over dt} P_O(t) &=& \Gamma_0 
e^{- \beta \epsilon_O} V_O \sum_{\gamma \ne O} V_{\gamma}
P_{\gamma}(t) \\
{d \over dt} P_{\alpha}(t) &=& \Gamma_0 \left( e^{-\beta
\epsilon_{\alpha}} V_{\alpha} \sum_{\gamma \ne O} V_{\gamma}
P_{\gamma}(t) -  V_{\alpha} P_{\alpha}(t) 
\sum_{\gamma}  e^{-\beta \epsilon_{\gamma}} V_{\gamma} \right)
\eea

These equations can be solved along the lines described above. The
Laplace transform of the occupation probability of the native state
is:
\be
\label{PO}
\tilde{P}_O(z) = {P_O(0) \over z} + \Gamma_0 {e^{-\beta \epsilon_O} V_O \over z} \tilde{Q}(z)
\ee
where 
\bea
\tilde{Q}(z) &=& { \sum_{\alpha \ne O} {P_{\alpha}(0) V_{\alpha} \over
z+\Gamma_0 V_{\alpha} \zeta } \over  1 - \Gamma_0  \sum_{\alpha \ne O}
{e^{-\beta \epsilon_{\alpha}} V_{\alpha}^2 \over z+\Gamma_0 V_{\alpha}
\zeta}}\\
\zeta &=& \sum_{\gamma} e^{-\beta \epsilon_{\gamma}} V_{\gamma} 
\eea
and the $P_{\alpha}(0)$ are the initial probabilities.

As in the previous section, the large time behavior of $P_O(t)$ is
governed by the pole structure of (\ref{PO}). As usually, there is a
pole at $z_p=0$ corresponding to the static limit. It is easily seen
on (\ref{PO}),
that the residue of $\tilde{P_O}(0)$ at $z=0$ is equal to 1. This
implies that:
\be
\lim_{t \to \infty} P_O(t) = 1
\ee
Contrarily to thermal equilibrium, the equilibrium solution populates
entirely the native state. 

The rest of the  poles are given
by the solutions of the equation:
\be
\label{isol}
  \Gamma_0  \sum_{\alpha \ne O}
{e^{-\beta \epsilon_{\alpha}} V_{\alpha}^2 \over z+\Gamma_0 V_{\alpha}
\zeta} = 1
\ee
This equation can again be solved graphically.
By
contrast to (\ref{Ptilde}),  where the sum runs over all states, here
it runs over all states except the native one. As a result, $z=0$ is
not anymore a solution of this equation; it is however a pole
of $\tilde{P}_{\alpha}(z) \ $(see above) . 

To be more specific and simplify the calculations, 
we discuss the case analogous to model I above.
We take
$V_{\alpha} = V$ . Equations (\ref{Pt}) can be solved
explicitly. Assuming that the native state is not populated at $t=0$,
we find:
\be
P_O(t) = 1 - e^{-\Gamma_0 V^2 e^{-\beta \epsilon_O} t}
\ee
and the quenched average can be performed, yielding:
\be
\overline{P_O(t)} \sim_{t \to \infty} 1 - {C\over t}\  e^{-\Gamma_0 V_0^2 e^{-\beta
\epsilon^*}t}
\ee
where $C$ is a constant.

We see on this very simple model that the introduction of a trap 
(in the energy landscape) from
which the system cannot escape, modifies the relaxations from stretched
exponential or algebraic decay, to simple exponential. In addition,
the yield of the relaxation is 1, instead of a Boltzmann weight. This
feature is quite independent on the specific form of the transition
rates used.

This qualitative feature is observed in experiments, where
the introduction of chaperones is seen to enhance the folding rates by
several order of magnitudes, and to increase significantly the yield
of 
folding \cite{Todd}

\subsection{Conclusion}
\label{sec:conc}
We have proposed two methods to approach the dynamics of protein
folding. Numerical simulations seem to support a two timescale picture for
protein folding \cite{Shakh_Gut1,Ca_Thi}: i) a fast hydrophobic driven collapse followed by ii)
a slow rearrangement process in the compact globule.

Our first approach, which strongly relies on the polymeric
character of the protein, yields results which should be valid in the
short time limit, and thus applicable to stage i). On the other hand, the phenomenological approach
stresses the heterogeneity of the system and describes the transition between
the various compact conformations. It should therefore be applicable
to stage ii). 

\section{General conclusion}
We have presented a disorder oriented view of the protein folding
problem. A protein is modelled by a heteropolymeric chain, with
several possibles types of interactions among the monomers. 
Both static
and dynamical issues were tackled. These models, as well as our 
approximate treatment thereof, may be a first step towards a more
realistic description of proteins. Among the weaknesses of the
approaches we have presented, let us mention the schematic interactions,
the thermodynamic limit,
the continuous description, the quenched average ( instead
of well defined sequences), the mean-field and ground-state dominance
approximation, etc...These approximations are used to render
calculations tractable. Although they might sometimes be crude and
irrealistic, 
we believe that our
approach is useful, since it may shed some light on the 
connection of the folded
state with other glassy systems.

\section*{Appendix A: Basic polymer physics}
\label{A}
\par
\subsection*{A1: General Introduction}
\label{general}
In this section, we introduce simple concepts of polymer
physics that will be used in the following sections. General
references, with a larger scope, include the books by
de Gennes \cite{PGG}, Doi and Edwards \cite{DOI}, des Cloizeaux
and Jannink \cite{descl}, and Freed \cite{Kaf}.\par

A polymer chain in a solvent is modelled as a sequence of $N$ links
representing the  monomers. Typical values of the  polymerization index $N$  
are $10^5$ in polymer chemistry, $10^2 - 10^3$ in proteins, and up to 
$10^{10}$ for DNA molecules, which are the longest known polymers.
Homopolymers are built from identical monomers, whereas heteropolymers
may contain different units (e.g. $20$ aminoacids for proteins or $4$
nucleotides for DNA).

In the following, we will consider only the single chain
problem; indeed, the (in vitro) 
protein folding problem is a priori a one chain
problem (we leave aside the question of chaperone assisted folding).
The partition function for the chain can be written as:
\begin{equation}
Z = \int \prod_{i} {d} {\vec r_i} \prod_{i} 
g({\vec r_i},{\vec r_{i+1}}) \ \exp ( - \beta {\cal H} )
\label{append1}
\end{equation}
where $\beta$ is the inverse of the temperature $T$.
The vectors ${\vec r_i}$ denote the positions of the nodes. For
example, in polyethylene, the ${\vec r_i}$ would represent
the positions of the carbon atoms. For future use, we will assume that
the polymer is embedded in a $d-$dimensional space.

The chain constraint is expressed by the factors $g(\vec r, {\vec
r'})$. Possible choices can be for instance:\par
(i) $ g(\vec r, {\vec r'}) = \delta ( |\vec r - {\vec r'}| - a ) $ ,
for freely hinged monomers of fixed length $a$.\par
(ii) one can further modify $g$ by restricting the nodes
to belong to a regular lattice, which is a quite common simplification
in Statistical Mechanics.\par
(iii) $ g(\vec r, {\vec r'}) = 
({d \over 2 \pi a^2})^{{d \over 2}}\ \exp (- d \ {( \vec r - {\vec r'}
)^2 \over 2 a^2 } ) $ , if the monomer length is allowed to fluctuate around $a$.

If one wishes to include curvature (resp. torsion ) effects, one
should use a function $ g(\vec r_i, {\vec r_{i+1}},{\vec r_{i+2}} ) $ (
resp. $ g(\vec r_i, {\vec r_{i+1}}, {\vec r_{i+2}}, {\vec r_{i+3}} ) $ ).

The ``Hamiltonian'' ${\beta \cal H}$ may be expanded as:
\begin{equation}
\label{append2}
{\beta \cal H} = {1 \over 2} \sum_{i \ne j} v( \vec r_i, \vec r_j) + { 1 \over
6 } \sum_{i \ne j \ne k} w( \vec r_i, \vec r_j, \vec r_k) + ...
\end{equation}
and represents the direct monomer-monomer 
interactions (such as Coulomb or Van der Waals interactions)
as well as the solvent-induced interactions. The particular case of a
chain with no interactions (${\cal H}=0$) is called a Brownian chain.
\par
In many polymer applications, one is interested in quantities which
do not depend on the  microscopic scale, hence the need for a continuum
description. Loosely speaking, this continuous limit is obtained (in
a field theoretic approach) by taking
$N \rightarrow +\infty $ and $a \rightarrow 0$
in such a way that the product $ S = N a^2 $, named the Brownian
surface of the chain, remains constant. More rigorous arguments can be
found in the book by des Cloizeaux
and Jannink \cite{descl}.
\par 
From a practical point of view, it is sufficient to replace everywhere
the discrete indices $i , j , k , \ldots  =1,\ldots,N $  of eq. (\ref{append1})
and (\ref{append2}) by continuous curvilinear abscissae $ s , s', s'', \ldots $.
Neglecting curvature and torsion effects, the partition function of
the chain can be written as:
\begin{equation}
\label{append3}
Z = \int {\cal D} {\vec r(s)} \exp ( - {d \over 2 a^2 }\int_0^N ds
\left( {d {\vec r(s)} \over ds} \right)^2 - \beta{\cal H} )
\end{equation}
with
\begin{equation}
\label{append4}
\beta {\cal H} = {1 \over 2}\int_0^N ds \int_0^N ds' v({\vec r(s)},{\vec
r(s')}) + {1\over 6} \int_0^N ds \int_0^N ds'\int_0^N ds'' w({\vec r(s)},{\vec
r(s')},{\vec r(s'')})
\end{equation}
\par
In the following, we shall use indistinctively the continuous or
discrete versions of the partition function. However, as we shall see
in appendix B,
the continuous  version is easier to handle for
analytic calculations, since it can be mapped onto an
imaginary time Schr\"odinger equation.\par
\subsection*{A2: Solvent-induced interactions}
\par
It is physically intuitive that, in a good solvent, a polymer chain
will swell so as to maximize contacts between monomers and solvent
molecules. On the contrary, a bad solvent will lead to a collapse of
the chain. One may equally well say that a good (resp. bad) solvent generates 
an effective repulsive (resp. attractive) monomer-monomer interaction. In the
context of protein folding, it is widely believed that the
interactions between the
hydrophobic
residues (Trp, Ile, Phe, ...) are the main driving forces for the collapse
of proteins: hydrophobic residues in water can be thought of as being
in a bad solvent, and thus have an attractive effective interaction.
\par
Let us consider a lattice model,
with $N$ monomers ${\vec r_i}$ and $\cal N$
solvent molecules ${\vec R_{\alpha}}$,  where each site is occupied
either by a monomer or a solvent
molecule or a vacancy.
The Hamiltonian of this model reads: 
\begin{equation}
\label{append5}
{\cal H}_{ms} = \sum_{i=1}^{N} \sum_{\alpha=1}^{{\cal N}} a_0 ( {\vec
r_i}-{\vec R_{\alpha}} ) 
\end{equation}
where $a_0$ denotes a short-ranged monomer-solvent molecule interaction. 
\par
Introducing the monomer and solvent volume concentrations:
\begin{equation}
\label{append6}
\rho_m(\vec r) = \sum_{i=1}^{N} \delta( {\vec r} - {\vec r_i} )
\end{equation}
and
\begin{equation}
\label{append7}
\rho_s(\vec r) = \sum_{\alpha=1}^{\cal N} \delta( {\vec r} - {\vec
R_{\alpha}} )
\end{equation} 
the Hamitonian (\ref {append5}) can be rewritten as:
\begin{equation}
\label{append8}
{\cal H}_{ms} = \sum_{\vec r} \sum_{{\vec r'}} \rho_m(\vec r) a_0 ( {\vec r}
- {\vec r'} ) \rho_s(\vec r')
\end{equation}
\par
Assuming that the system is incompressible (no vacancies), we have, at each
site $\vec r$ :
\begin{equation}
\label{append9}
\rho_s(\vec r)+\rho_m(\vec r) = 1
\end{equation}
Replacing (\ref {append9}) in (\ref {append8}) we obtain:
\begin{equation}
\label{append10}
{\cal H}_{ms} = - \sum_{\vec r} \sum_{{\vec r'}} \rho_m(\vec r) a_0 ( {\vec r}
- {\vec r'} ) \rho_m(\vec r') +  \sum_{\vec r} \sum_{{\vec r'}}
 \rho_m(\vec r) a_0 ( {\vec r} - {\vec r'} ) 
\end{equation}

The second term in (\ref {append10}) is a constant, equal to $ N A_0 $ where by definition
$A_0 = \sum_{\vec r} a_0 ( {\vec r} )  $. It will therefore be omitted.
\par
Note the change of sign of the first term between (\ref {append8}) and
(\ref {append10}). It
generates, as announced above, a contribution to the two-body
interaction $v({\vec r}- {\vec r'})$ of (\ref {append2}). The good (resp. bad)
solvent is characterized by a negative (resp. positive) $ a_0 ( {\vec
r} )$, and indeed generates a repulsive (resp. attractive) interaction    
$- a_0 ( {\vec r} )$ between monomers.
In order to apply these considerations to heteropolymers or proteins,
it is necessary to allow for sequence dependent interactions $a_0$ in
equation (\ref {append5}). This will be discussed in section \ref{sec:hydro}.\par

\section*{Appendix B: Some methods of polymer physics}
\label{Bibi}
In this appendix, we review the properties of free Brownian chains,
and show their relation to diffusion and Schr\"odinger equations. For
interacting chains, we present
the self-consistent field approximation (SCF), 
and show how it can be further simplified when there is a gap in the energy
spectrum, by using the ground state dominance approximation.

\subsection*{B1: Markovian and Brownian chains.}

Let us consider a discrete chain, made of $N$ monomers; the first
atom of the chain, numbered $0$, is fixed at point $\vec 0$. By definition
of a Markov chain, the position $\vec r_n$ of atom $n$ depends only
on the position $\vec r_{n-1}$ of the previous atom $n-1$ along the
chain. Namely, a Markov chain is characterized by the conditional
probability distribution $g(\vec r,\vec r')$ that atom $n$ is at
point $\vec r$, given that atom $n-1$ is at $\vec r'$. If we further
assume translational invariance, this function will only depend on the
difference, and reads $g(\vec r-\vec r')$.
\par
The probability of finding the end of the chain (atom number $N$) at
point $\vec R$, given that the origin is fixed at $\vec 0$, is thus given by:
\begin{equation}
\label{green1}
G(\vec R,N|\vec 0,0) = \int {d}^d r_1 \ldots { d}^d r_{N-1} 
\ g(\vec R -\vec r_{N-1})\ g(\vec r_{N-1} -\vec r_{N-2}) \ldots g(\vec
r_1 - \vec 0)
\end{equation}
The Fourier transform $\Gamma$ of $G$ is thus given by:
\begin{equation}
\label{green2}
\Gamma (\vec k,N) = \int {d}^d R \ e^{i \vec k \vec R}\ G(\vec
R,N|\vec 0,0) = \gamma^{N}(\vec k) 
\end{equation}
where
\begin{equation}
\label{green3}
\gamma(\vec k)= \int {d}^d r \ e^{i \vec k \vec r}\ g(\vec r)
\end{equation}
Performing an inverse Fourier transform, we obtain:
\begin{equation}
\label{green4}
G(\vec R,N|\vec 0,0) = \int {d^d k \over (2\pi)^d}
e^{-i \vec k \vec R}\ \gamma^{N} (\vec k)
\end{equation}
In the limit of long chains, $N \to \infty$, we apply the
stationary-phase method to (\ref {green4}). Since $g(\vec r)$ is a
probability distribution, its integral is one, and thus it is easy to
see that:
\begin{equation}
\label{green5}
|\gamma (\vec k)| \le \gamma (\vec 0) =1
\end{equation}
so that we may expand (\ref {green4}) around $\vec k = \vec 0$. 
We can write, to second order in $\vec k$:
\begin{equation}
\label{green55}
\gamma(\vec k) = 1 - {a^2 \over 2d} {\vec k}^2
\end{equation}
where $a$, which has the dimension of a length, is called the
Kuhn length, and can be interpreted as the effective monomer length.
We have:
\begin{equation}
\label{green6}
G(\vec R,N|\vec 0,0) = \int {{d}^d k \over (2\pi)^d}
\exp(-i \vec k \vec R\ -N {a^2 \over 2d} {\vec k}^2)
= \left( {d \over 2 \pi N a^2 } \right) ^{d/2} e^{-{d R^2 \over 2 N
a^2}}\nonumber\\
\end{equation}
We see that, for long enough chains, regardless of the function $g(\vec r)$,
the asymptotic probability distribution for the end of the chain is
Gaussian. As an outcome, we see that the entropy (more precisely the
entropy reduction) of a chain, with
origin constrained at $\vec 0$ and extremity constrained at $\vec
R$ is given by:
\begin{equation}
\label{green7}
S = - { d R^2 \over 2 N a^2}
\end{equation}
a result that we shall frequently use.
\par
The expression (\ref {green7}) of $G(\vec R,N|\vec 0,0)$ shows that it
satisfies a diffusion-like equation:
\begin{equation}
\label{green8}
\left({\partial \over \partial N} -{a^2 \over 2 d } {\vec \nabla}^2\right) G(\vec
R,N|\vec 0,0) = \delta(\vec R) \delta(N)
\end{equation}
which is nothing but an imaginary-time Schr\"odinger equation.
Therefore, using the standard notations of quantum mechanics, we see
that we can write:
\begin{equation}
\label{green9}
G(\vec R,N|\vec 0,0) = <\vec R| e^{ - N {\cal H} } | \vec 0>
\end{equation}
where $\cal H$ is a quantum-like Hamiltonian equal to:
\begin{equation}
\label{green10}
{\cal H} =- {a^2 \over 2 d } {\vec \nabla}^2
\end{equation}
It is well-known in quantum mechanics \cite{Neg_O},
that matrix elements of the form (\ref {green9}), have a path-integral
representation. We may thus write:
\begin{equation}
\label{blue1}
 G(\vec R,N | \vec 0,0) = \int_{\vec r(0) =\vec 0}^{\vec r(N) =\vec R}
 {\cal D} {\vec r}(s) \exp\left(-{d \over 2 a^2} \int_0^N { d}s \ \dot {
\vec r}^2(s)\right) 
\end{equation}
This continuous chain is called a Brownian chain. It is the prototype
of all long chains defined by Markov processes.
\par
These expressions can easily be generalized to the case of a chain in
an external potential. Assuming that each atom $\vec r_n$ feels the
potential $U(\vec r_n)$, the distribution function of the chain at inverse
temperature $\beta$ reads:
\begin{equation}
\label{blue2}
 G(\vec R,N | \vec 0,0) = \int_{\vec r(0) =\vec 0}^{\vec r(N) =\vec R}
 {\cal D} {\vec r}(s) \exp\left(-{d \over 2 a^2} \int_0^N { d}s \ \dot{
\vec r}^2(s) - \beta \int_0^N { d}s\ U(\vec r(s))\right) 
\end{equation}
It has a form similar to (\ref {green9}) with
\begin{equation}
\label{blue3}
{\cal H} =- {a^2 \over 2 d } {\vec \nabla}^2 + \beta \ U(\vec r)
\end{equation}
and satisfies the Schr\"odinger equation:
\begin{equation}
\label{blue4}
\left({\partial \over \partial N} -{a^2 \over 2 d } {\vec \nabla}^2 +
 \beta \ U(\vec r) \right) G(\vec
R,N|\vec 0,0) = \delta(\vec R) \delta(N)
\end{equation}
The solution of (\ref {blue4}) can easily be expressed in terms of the
eigenstates and eigenvalues of the quantum Hamiltonian $\cal H$.
Denoting by $\Psi_{n} (\vec r)$ the eigenfunction of $\cal H$ with
eigenvalue $E_n$, which satisfy the equation : 
\begin{equation}
\label{blue5}
{\cal H} \Psi_{n} (\vec r) = E_{n} \Psi_{n} (\vec r)
\end{equation}
we can rewrite:
\begin{equation}
\label{blue6}
G(\vec R,N|\vec 0,0) = \sum_{\{n\}} e^{-N E_{n} } \Psi_{n}(\vec R)
\Psi_{n}(\vec 0)
\end{equation}

\subsection*{B2: Ground state dominance.}
\label{ground}
An important simplification occurs when:\par
(i) the length $N$ goes to $\infty$.\par
(ii) there is a gap $\Delta$ in the energy spectrum of $\cal H$, such
that $(N\Delta \gg 1)$.\par
(iii) there is not an exponentially large number of excited states in
this spectrum.\par 
Indeed, in this case, the sum in equation (\ref{blue6}) is dominated by
the ground state $\Psi_0$ with energy $E_0$, and the distribution function
reduces to:
\begin{equation}
\label{blue7}
G(\vec R,N|\vec 0,0) \sim  e^{-N E_{0} } \Psi_{0}(\vec R)
\Psi_{0}(\vec 0)
\end{equation}
Condition (i) and (ii) implies that, for a long enough chain, the
Hamiltonian $\cal H$ has bound states. This is the case for the
adsorption of a polymer chain on an impenetrable wall.
\par
If condition (iii) is not satisfied, or if the spectrum is continuous with
no finite gap, one has to solve the
full ``time''-dependent Schr\"odinger equation.
\par
To conclude this section, let us recall that the ground state of a
Hamiltonian can be deduced from a variational principle, namely the
Rayleigh-Ritz principle. This principle states that the ground state
energy $E_0$ of a Hamiltonian $\cal H$ is given by:
\begin{equation}
\label{blue8}
E_0 = \min_{\{\Psi(\vec r)\}} {<\Psi|{\cal H}|\Psi> \over <\Psi|\Psi>}
\end{equation}
where the minimization runs over the space of square-integrable functions.
Using the energy as a Lagrange multiplier to enforce normalization, it
is equivalent to minimize the functional $\cal E$:
\begin{equation}
\label{blue9}
{\cal E} = \int {d}^d r
\ \Psi (\vec r) \left(-{a^2 \over 2 d } {\vec \nabla}^2 +  \beta \ U(\vec r)
 - E_0\right) \Psi (\vec r) 
\end{equation}
\subsection*{B3: Self consistent field approximation.}
\par
The previous sections have shown how one can generate path integral
representations for polymer problems. In the case of a
self-interacting chain, however, there is no equivalent to the
Schr\"odinger equation, and one has to resort to approximations to
solve the problem. One such powerful approximation was devised by
Edwards \cite{Edw} and is called the self-consistent field
approximation (SCF). We first illustrate the method by the case of a
chain in a bad solvent.
\subsubsection*{SCF with ground state dominance}
\label{SCFgsd}
The partition function of this model reads:
\bea
\label{red1}
Z =& \int{\cal D}\vec  r(s)\ \exp 
\left( -{d \over
2 a^2} \int^ N_0 { d} s\ \dot{\vec r}^2 \right)
 \ \exp \left( {v \over 2} \int^ N_0 { d} s { d}
s^{\prime} \delta \left(\vec  r(s)-\vec  r \left(s^{\prime} \right) 
\right)\right)\nonumber\\
& \times \exp\left(- {w \over 6}  \int^ N_0 { d} s \ { d} s^{\prime}
\ { d} s^{\prime \prime}
\delta \left(\vec  r(s)-\vec  r \left(s^{\prime} \right) \right)
\delta \left(\vec  r(s)-\vec  r \left(s^{\prime \prime} \right) \right)
\right)
\eea
It is very useful to make a change of variable on (\ref {red1}), so that
the partition function of the problem is expressed as a functional
integral over all possible monomer concentration $\{\rho(\vec r)\}$. To
do so, we enforce the variables $\{\rho(\vec r)\}$ by inserting the
identity:
\begin{equation}
\label{red2}
1= \int {\cal D} \phi ( \vec r ){\cal D} \rho ( \vec r )
\exp \left ( i \int d^d r \phi ( \vec r ) \rho ( \vec r )
-i \int^ N_0 { d} s\ \phi \left(\vec  r(s) \right) \right)
\end{equation}
which expresses that:
\begin{equation}
\label{red3}
\rho (\vec r) = \int_0^N { d} s \ \delta( \vec r - \vec r(s))
\end{equation}
Inserting these identities in (\ref {blue8}) yields:
\begin{equation}
\label{red4}
Z = \int {\cal D} \phi ( \vec r ){\cal D} \rho ( \vec r )
\exp \left ( i \int d^d r \phi ( \vec r ) \rho ( \vec r ) 
+ {v \over 2} \int d^d r \rho^2 (\vec r) 
- {w \over 6} \int d^d r \rho^3 (\vec r) \right ) \zeta (\phi)
\end{equation}
where 
\begin{equation}
\label{red5}
\zeta (\phi) =
\int^{ }_{}{\cal D}\vec  r(s)\ 
{\exp} \left(-{d \over 2a^2} \int^ N_0 { d} s\ {\dot  {\vec r}^2}-i \int^ N_0 { d} s\ \phi
\left(\vec  r(s) \right) \right) 
\end{equation}
According to (\ref {blue2}), $\zeta$ can be expressed as:
\begin{equation}
\label{red6}
\zeta (\phi) = \int { d}^d r  <\vec r| e^{-N{\cal H}}|\vec 0>
\ee
where
\be
\label{red7}
{\cal H} = -{a^2 \over 2d} {\vec \nabla}^2 + i \phi(\vec r)
\ee
In a bad solvent, we expect the chain to collapse at some temperature,
and therefore, there should be some bound state in the system. We thus
assume ground state dominance, so that:
\be
\label{red8}
\zeta (\phi) = \Psi_0(\vec 0) \left( \int { d}^d r \Psi_0(\vec r) \right) 
\exp(- N E_0) 
\ee
where $\Psi_0$ is the ground state of $\cal H$, with energy $E_0$.
Using equations (\ref {blue8}), we have seen that the extensive part
of $\zeta$ can be written as:
\be
\label{red9}
\zeta = \exp \left( - N \min_{\{\Psi(\vec r)\}} {<\Psi|{\cal H}|\Psi>
\over <\Psi|\Psi>} \right) 
\ee
or equivalently
\be
\label{red10}
\zeta = \exp \left(-N \min_{\{\Psi(\vec r)\}}\left[  \int { d}^d r
\ \Psi (\vec r) \left(-{a^2 \over 2 d } {\vec \nabla}^2 +  i \phi(\vec r)
 \right) \Psi (\vec r) - E_0 \left( \int { d}^d r
\ \Psi^2 (\vec r) -1  \right) \right] \right) 
\ee
where $E_0$ appears as a Lagrange multiplier which constrains the norm
of $\Psi^2$ to 1.
\par
Replacing (\ref {red10}) in (\ref {red4}) yields:
\be
\label{white1}
Z = \int {\cal D} \phi ( \vec r ){\cal D} \rho ( \vec r )
\exp \left( -N {\cal F}(\rho,\phi) \right)
\ee
where
\bea
\label{white2}
N {\cal F}(\rho,\phi) = - i \int d^d r \phi ( \vec r ) 
\rho (\vec r)
- {v \over 2} \int d^d r \rho^2 (\vec r) 
+ {w \over 6} \int d^d r \rho^3 (\vec r)\nonumber\\
+  N \min_{\{\Psi_0(\vec r)\}} \left[  \int { d}^d r 
\ \Psi (\vec r) \left(-{a^2 \over 2 d } {\vec \nabla}^2 +  i \phi(\vec r)
 \right) \Psi (\vec r) - E_0 \left( \int { d}^d r
\ \Psi^2 (\vec r) -1  \right) \right] 
\eea
Since (\ref {white1}) cannot be evaluated exactly, we must use some
approximation. A natural approximation is the saddle-point method
(SPM), which consists in expanding ${\cal F}(\rho,\phi)$ around its
minimum.
\par
The minimization equations with respect to $\phi(\vec r)$, $\rho(\vec
r)$ and $\Psi (\vec r)$ read:
\bea
\label{white3}
\rho(\vec r) =&& N \Psi^2 (\vec r)\nonumber\\
i \phi(\vec r) =&& -v \rho(\vec r) +{w \over 2} \rho^2(\vec r)\nonumber\\
 E_0 \Psi(\vec r)=&& \left(-{a^2 \over 2 d } {\vec \nabla}^2 + i \phi(\vec r) \right)
\Psi(\vec r) 
\eea
The first equation expresses that the monomer concentration is $N$
times the square of the normalized wave-function. The second equation
expresses the mean-field potential seen by each monomer, and finally
the last equation can be recast in the form:
\be
\label{white4}
\left(-{a^2 \over 2 d} {\vec \nabla}^2 -v N \Psi^2 (\vec r) +{w \over 2} 
N^2 \Psi^4(\vec r) \right) \Psi(\vec r) = E_0 \Psi(\vec r)
\ee
where the energy $E_0$ is chosen so that the square wave function is
normalized. The above equation is a non-linear Schr\"odinger equation, which
cannot in general be solved analytically. Let us note however that the
two-body term $v$ plays the role of an attractive self-consistent
field, whereas $w$ plays the role of a repulsive one, which forbids
the collapse onto a finite region. In dimension larger than 2, this
equation will have a bound state if $v$ is large enough, and the
associated energy will be finite, even in the limit $N \to \infty$.
\par
To get more analytic information, we can restrict the space of
normalized wave functions $\{\Psi(\vec r)\}$ to the set of Gaussian
wave-functions, depending on a single parameter $R$, which measures the
spatial extent of the polymer globule:
\be
\label{white5}
\Psi(\vec r) =  \left( {1\over 2 \pi R^2} \right)^{d/4}
\exp(- { {\vec r}^2 \over 4 R^2})
\ee
Replacing (\ref {white5}) in (\ref {white2}) yields the simple
equation for $R$: 
\be
\label{white6}
0={a_0 \over R^3 } - v a_2 {N \over R^{d+1}} +w a_3 {N^2 \over
R^{2d+1}}
\ee
where $a_0$, $a_2$ and $a_3$ are simple numerical constants. 
This equation shows that for large $N$ and any attractive two-body
interaction, the system is collapsed in a globular state, with finite
concentration, with exponent $\nu = 1/d$. Note that this implies that
the kinetic energy term  ${Na_0\over R^2}$ yields (for $d>2$) a 
vanishing contribution in the collapsed phase. Since this term is directly
linked to the chain constraint (it is in fact the entropy loss due to
the collapse transition) , we can see that the above treatment is
not fully satisfactory. It does not describe well the extensive
conformational entropy of the collapsed phase \cite{Orla_dd}.

\par
\subsubsection*{SCF without ground state dominance}
\label{SCG}
We now consider the case where the ground state dominance
approximation is not valid. Note that this may arise in (at least) two
ways: either one may 
have to deal with a continuum spectrum for the Hamiltonian previously
defined, or one may be faced with an exponentially large number of
metastable states above the ground state. The former possibility is
met in the self-avoiding chain, the latter has not been encountered
yet, but should be present in some heteropolymer problems (one may have a
``non-zero complexity'', as in the p-spin glass models $p\ge3$). An
example of such a behavior is probably the coil phase ``with
metastable states'' of section \ref{sec:hydro}.
For the sake of simplicity, we only include two-body interactions in
the Hamiltonian.
In that case, one cannot use (\ref {red8}). One can still write saddle point
equations for the variables $\phi(\vec r)$ and $\rho(\vec r)$ in
equation (\ref{red4}). These equations in turn yield a self consistent
equation for $\rho(\vec r)$. We have chosen here a slightly different
presentation, to establish a possible connection with the spin glass
TAP equations. The partition function reads
\bea
\label{ed1}
Z = &\int{\cal D}\vec  r(s)\ \exp 
\left( -{d \over
2 a^2} \int^ N_0 { d} s\ {\dot  {\vec r}}^2 
-{1 \over 2} \ \int^ N_0 { d} s \int^ N_0 { d}
s^{\prime} \ v(s,s^{\prime}) \ \delta \left(\vec  r(s)-\vec  r (s^{\prime}) \right) 
\right)
\eea
and may be rewritten through a Hubbard Stratanovich transformation
\bea
\label{ed2}
Z =&&\int{\cal D}\Phi(\vec r,s) \ \exp\left( -{1 \over 2} \int d^d r \ \int^ N_0 { d} s \int^ N_0 { d}
s^{\prime} \ \Phi(\vec r,s) {v(s,s^{\prime})}^{-1} \Phi(\vec
r,s^{\prime}) \right)\nonumber \\
&& \int{\cal D}\vec  r(s) \exp \left( -{d \over 2 a^2} \int^ N_0 { d}
s\ {\dot  {\vec r}}^2 \right) \ \exp\left(-i \int^ N_0 { d} s \Phi(\vec
r(s),s) \right)   
\eea
The polymer integral may be rewritten as a Feynman path integral
analogous to (\ref {blue1}), with both ends free. We have
\bea
\label{ed3}
Z &=& \int{\cal D}\Phi(\vec r,s) \ \exp\left( -{1 \over 2}\int d^d
r \ \int^ N_0 { d} s \int^ N_0 {d}
s^{\prime} \ \Phi(\vec r,s) v^{-1}(s,s^{\prime}) \Phi(\vec
r,s^{\prime}) \right) \nonumber \\ 
&& \times \int d^d r  \ \int d^d r^{\prime} \ G(\vec r N\vert
\vec r^{\prime}0)
\eea
where the matrix element $G(\vec r,N \vert\vec r^{\prime},0)$
satisfies the equation 
\be
\label{ed4}
\left({\partial \over \partial N} -{a^2 \over 2 d } {\vec \nabla}^2 +
 i\Phi(\vec r,N) \right) G(\vec
r,N|\vec r^{\prime},0) = \delta(\vec r-\vec r^{\prime}) \delta(N)
\end{equation}
Due to the first order character of this equation with respect to the
``time'' $N$, one may rewrite the matrix element $G$ as
\bea
\label{ed5}
G(\vec r,N\vert\vec r^{\prime},0)=&& \int{\cal D} \Psi(\vec y,s) \int{\cal
D} \Psi^{\dagger}(\vec y,s)  \Psi(\vec r,N) \ \Psi^{\dagger}(\vec
r^{\prime},0) \nonumber\\
&&\times \exp\left( - \int d^d {\rho} \int^ N_0 { d}
s \Psi^{\dagger}(\vec {\rho},s)\left({\partial \over \partial s}+{{\vec
p }^{2}\over 2}+i\Phi(\vec {\rho},s)\right) \Psi(\vec {\rho},s) \right) 
\eea
where ${\vec p}^{2}=-{a^2\over {2d}} {\vec \nabla}^2$. Plugging back equation (\ref{ed5}) 
in (\ref{ed3}), one may now
integrate over the $\Phi(\vec {\rho},s)$.We obtain
\bea
\label{ed6}
Z&=&\int{\cal D} \Psi(\vec y,s) \int{\cal
D} \Psi^{\dagger}(\vec y,s)  \Psi(\vec r,N) \ \Psi^{\dagger}(\vec
r^{\prime},0) \nonumber\\
& & \times \exp\left( - \int d^d {\rho} \int^ N_0 { d}
s \Psi^{\dagger}(\vec {\rho},s)\left({\partial \over \partial
s}+{{\vec
p }^{2}\over 2}\right) \Psi(\vec {\rho},s) \right) \nonumber\\
& &\times \exp\left( -{1 \over 2} \int  d^d
{\rho} \ \int^ N_0 { d} s \int^ N_0 { d}
s^{\prime} \ \vert{\Psi(\vec {\rho},s)}\vert^{2} {v(s,s^{\prime})}
\vert{\Psi(\vec {\rho},s^{\prime})}\vert^{2} \right)  
\eea
where the short hand notation $\vert{\Psi(\vec
{\rho},s)}\vert^{2}=\Psi(\vec {\rho},s)\Psi^{\dagger}(\vec {\rho},s)$
was used.  
We now may performed the saddle point method with respect to both
$\Psi$ and $\Psi^{\dagger}$. We have for instance
\be
\label{ed7}
\left({\partial \over \partial s}-{a^2\over {2d}}{\vec \nabla}^{2}+\int { d}
s^{\prime} {v(s,s^{\prime})} \vert{\Psi(\vec
{\rho},s^{\prime})}\vert^{2}\right) \Psi(\vec {\rho},s)=0
\ee
with the boundary conditions $\Psi(\vec r,0)=\delta(\vec r)$, and a similar
equation for $\Psi^{\dagger}$.\par
The adjunction of three (and more) body interactions is
straightforward. In principle, these equations may be solved. Edwards
has studied the case of the self avoiding chain \cite{Edw} and
obtained the Flory value of the swelling exponent $\nu=3/5$. More
recently, this method has been used recently \cite{Ma_Sch} to find the
phase diagram of (non random) block copolymer melts. In the presence
of disorder, these equations are certainly difficult to solve: they
are the equivalent of the spin glass TAP equations, which have 
not been solved numerically so far.


\section*{References}
\begin{thebibliography}{99} 

\bibitem{Creighton84} T.E. Creighton,
{\em Proteins}, W.H. Freeman, New York (1984).
\bibitem{Creighton92} T.E. Creighton (editor),
{\em Protein Folding}, W.H. Freeman, New York (1992).
\bibitem{Dill95}
K.A. Dill, S. Bromberg, K. Yue, K.M. Fiebig, D.P. Yee, P.D. Thomas and
H.S. Chan, {\em Protein Science}, {\bf 4}, 561 (1995).
\bibitem{Elber96}
R. Elber, in {\em New Developments in
Theoretical Studies of Proteins}, R. Elber (ed.), World Scientific,
Singapore, (1996). 
\bibitem{Garel94}
T. Garel, H. Orland and D. Thirumalai, in {\em New Developments in
Theoretical Studies of Proteins}, R. Elber (ed.), World Scientific,
Singapore, (1996). 
\bibitem{Dar_Lod_Bal}
J. Darnell, H. Lodish and D. Baltimore, {\em Molecular Cell Biology}, 
Scientific American Books (1990).
\bibitem{Stryer}
L. Stryer, {\em Biochemistry}, W.H. Freeman, New-York (1988).
\bibitem{cell}
B. Alberts, D.  Bray, J. Lewis, M. Raff, K. Roberts  and J.D. Watson,
{\em Molecular Biology of the Cell}, Garland Publishing (1983)
 \bibitem{descl}
J. des Cloizeaux and G. Jannink, {\em Les Polym\`eres en 
Solution} Eds. de Physique, Les Ulis (France), 1987.
\bibitem{Pti}
O.B. Ptitsyn, {\em Adv. Protein Chem.}, {\bf 47}, 83 (1995).
\bibitem{pau_co}
L. Pauling and R.B. Corey R.B. {\sl
Proc.Natl.Acad.Sci.USA\/}, {\bf 37}, 235, 251, 272, 729
\bibitem{Me_Pa_Vi}
M. M\'ezard, G. Parisi and M.A. Virasoro, {\em Spin glass theory 
and beyond} , World Scientific, Singapore, (1987).
\bibitem{book}
see the other articles in this book.
\bibitem{Bin_You}
K. Binder and A.P. Young, {\em Rev. Mod. Phys.}, {\bf 58}, 801 (1986).
\bibitem{Brout}
R. Brout, {\em Phys. Rev.} , {\bf 115}, 824 (1959).
\bibitem{Aha_Har}
A. Aharony and A.B. Harris, {\em Phys. Rev. Lett.}, {\bf 77}, 3700 (1996).
\bibitem{Ki_Thi_Wo} 
T.R. Kirkpatrick, D. Thirumalai and P.G. Wolynes,  {\em Phys. Rev. A}, {\bf
40}, 1045 (1989).
\bibitem{Der} 
B.D. Derrida, {\em Phys. Rev. B}, {\bf 24}, 2613 (1981).
\bibitem{Gro_Som}
D.J. Gross, I. Kanter and H. Sompolinsky, {\em Phys. Rev. Lett.}, {\bf 55},
304 (1985).
\bibitem{IM}
Y. Imry and S.-k. Ma, {\em Phys. Rev. Lett.}, {\bf 35}, 1399 (1975).
\bibitem{Berk}
A.N. Berker, {\em Physica} A, {\bf 194}, 72 (1993).
\bibitem{Fr_Mi_Le}
G.H. Fredrickson, S.T. Milner and L. Leibler, {\em Macromolecules}, {\bf
25}, 6341 (1992). 
\bibitem{Obu} 
S.P. Obukhov, {\em J. Phys. A}, {\bf 19}, 3655 (1986).
\bibitem{Ga_Ga_O} 
J.R. Garel, T. Garel and H. Orland, {\em J. Phys. (France)}, {\bf 50}, 3067 (1989).
\par
\bibitem{Zim_Lev} 
B.H. Zimm and S.D. Levene, {\em Quart. Rev. Biophysics}, {\bf 25}, 171 (1992).
\bibitem{Ga_O}
T.  Garel, D.A. Huse, S. Leibler and H. Orland, {\em Europhys. Lett.}, {\bf 8},
9 (1989).
\bibitem{Sr_Ch_Sh}
S. Srebnick, A.K. Chakraborty and E.I. Shakhnovich,
{\em Phys. Rev. Lett.}, {\bf 77}, 3157 (1996).
\bibitem{Ed_Mu}
S.F. Edwards and M. Muthukumar, {\em J. Chem. Phys.}, {\bf 89}, 2435 (1988)
\bibitem{Ga_O_Le}
T. Garel, L. Leibler and H. Orland, {\em J. Phys. (France)} II, {\bf 4},
2139 (1994).
\bibitem{Mo_Ku_Da}
A. Moskalenko, Y.A. Kuznetsov and K.A. Dawson, {\em
J. Phys. (France)}, II, {\bf 7}, 409 (1997).  
\bibitem{Qia_Kho}
C. Qian and A.L. Kholodenko, {\em J. Chem. Phys.}, {\bf 89}, 5273 (1988).
\bibitem{Brazo}
S.A. Brazovskii, {\em JETP}, {\bf 41}, 85 (1975)
\bibitem{Al_Ta}
S. Alexander and J. McTague, {\em Phys. Rev. Lett.}, {\bf 41}, 702 (1978).
\bibitem{Me_Mo}
M. M\'ezard and R. Monasson, {\em Phys. Rev.} B, {\bf 50}, R7199 (1994).
\bibitem{Ca_Thi}
C.J. Camacho and D. Thirumalai, {\em Proc. Natl. Acad. Sci. USA}, {\bf 90},
6369 (1993).
\bibitem{Shakdyn}
E.I. Shakhnovich, {\em Phys. Rev. Lett.}, {\bf 72}, 3907 (1994).
\bibitem{Kli_Thi}
D.K. Klimov and D. Thirumalai, {\em Phys. Rev. Lett.}, {\bf 76}, 4070 (1996).
\bibitem{Gar_Orl1}
T. Garel and H. Orland,  
{\em Europhys. Lett.}, {\bf 6}, 307 (1988).
\bibitem{Bry_Woly}
J. Bryngelson and P.G. Wolynes, {\em Proc. Natl. Acad. Sci. USA}, {\bf 84},
7524 (1987).
\bibitem{Sh_Gu}
E.I. Shakhnovich and A.M. Gutin,  {\em J. Phys.} A, {\bf 22}, 1647 (1989).  
\bibitem{Ba_Bou_Me}
L. Balents, J-P. Bouchaud and M. M\'ezard, {\em J. Phys. (France) } I, {\bf
6}, 1007 (1996) and references therein.
\bibitem{Gar_Orl2} 
T. Garel and H. Orland,  
{\em Europhys. Lett.}, {\bf 6}, 597 (1988).
\bibitem{Sh_Gu2}
E.I. Shakhnovich and A.M. Gutin, 
{\em J. Phys. (France)}, {\bf 50}, 1843 (1989).
\bibitem{Lei}
L. Leibler, {\em Macromolecules}, {\bf 13}, 1602 (1980).
\bibitem{Dobry}
A.V. Dobrynin and I.Y. Erukhimovich, {\em J. Phys. (France)} I, {\bf 5},
677 (1995). 
\bibitem{Shak}
C. Sfatos, A.M. Gutin and E.I. Shakhnovich, {\em Phys. Rev.} E, {\bf 51},
4727 (1995).
\bibitem{Anger}
H. Angerman, G. ten Brinke and I. Erukhimovich, {\em Macromolecules}, {\bf
29}, 3255 (1996).
\bibitem{Ki_Thi}
T.R.  Kirkpatrick and D. Thirumalai, {\em J. Phys. A}, {\bf 22}, L149 (1989).
\bibitem {Imb}
J.M. Victor and J.B. Imbert, {\em  Europhys. Lett.}, {\bf 24}, 189 (1993).
\bibitem{Kard}
Y. Kantor and M. Kardar, {\em Europhys. Lett.}, {\bf 28}, 169 (1994).
\bibitem{Grass}
P. Grassberger and R. Hegger, {\em Europhys. Lett.}, {\bf 31}, 351 (1995).
\bibitem{der_hi}
B. Derrida, R.B. Griffiths and P.G. Higgs, {\em Europhys. Lett.}, {\bf 18},
361 (1992). 
\bibitem{Hi_Jo}
P.G. Higgs and J.F. Joanny, {\em J. Chem. Phys.}, {\bf 94}, 1543 (1991).
\bibitem{Gu_Sh}
A.M. Gutin and E.I. Shakhnovich, {\em Phys. Rev. E}, {\bf 50}, R3322 (1994).
\bibitem{Do_Ru}
A.V. Dobrynin and M. Rubinstein, {\em J. Phys. (France) II}, {\bf 5}, 677 (1995).
\bibitem{Pa_Gro}
V.S. Pande, A.Y. Grosberg, C. Joerg, M. Kardar and T. Tanaka,
{\em Phys. Rev. Lett.}, {\bf 76}, 3565 (1996).
\bibitem{Pan_Gro_Ta}
V.S. Pande, A.Y. Grosberg and T. Tanaka, {\em Proc. Natl. Acad. Sci. USA},
{\bf 91}, 12976 (1994).
\bibitem{Pa4}
V.S. Pande, A.Y. Grosberg, C. Joerg and T. Tanaka,
{\em Phys. Rev. Lett.}, {\bf 76}, 3987 (1996).
\bibitem{Plot_W}
S.S. Plotkin, J. Wang and P.G. Wolynes, {\em Phys. Rev. E}, {\bf 53}, 6271 (1996).
\bibitem{Ir_pet}
A. Irb\"ack, C. Peterson and F. Potthast, preprint chem-ph/ 9512004.
\bibitem{Frauen}
H. Frauenfelder, S.G. Sligar and P.G. Wolynes, {\em Science}, {\bf 254}, 
1598 (1991).
\bibitem{You_Pow}
R.D. Young and S.W. Powell, {\em J. Chem. Phys.}, {\bf 101}, 9919 (1994).
\bibitem{Eaton}
S.J. Hagen and W.A. Eaton, {\em J. Chem. Phys.}, {\bf 104}, 3395 (1996).
\bibitem{Angell}
J.L. Green, J. Fan and C.A. Angell, {\em J. Phys. Chem.}, {\bf 98}, 13780 (1994).
\bibitem{Deu_Ku}
J.M. Deutch and T. Kurosky, {\em Phys. Rev. Lett.}, {\bf 76}, 323 (1996).
\bibitem{Seno}
F. Seno, M. Vendruscolo, A. Maritan and J.R. Banavar, {\em
Phys. Rev. Lett.}, {\bf 77}, 1901 (1996).  
%
%
%
\bibitem{PGG1}  
P.G. de Gennes,    
{\it J. Physique Lett.},  {\bf 46}, L-639 (1985).

\bibitem{PGG2}  
A. Buguin, P.G. de Gennes and F.Brochart-Wyart,
{\it C.R.Acad. Sci. Paris}, {\bf 322}, 741 (1996).

\bibitem{BILL}  
C-K. Chan, Y. Hu, S. Takahashi, D.L. Rousseau, W. Eaton,
J. Hofrichter,    
{\it Proc. Natl. Acad. Sci. USA}, in press (1997).


\bibitem{TIM1}  
E.G. Timoshenko, K.A. Dawson,    
{\it Phys. Rev. E },  {\bf 51}, 492 (1995). 

E.G. Timoshenko, Yu.A. Kuznetsov, K.A. Dawson,    
{\it J. Chem. Phys.},  {\bf 102}, 1816 (1995). 

Yu.A. Kuznetsov, E.G. Timoshenko, K.A. Dawson,    
{\it J. Chem. Phys.},  {\bf 103}, 4807 (1995). 

Yu.A. Kuznetsov, E.G. Timoshenko, K.A. Dawson,    
{\it J. Chem. Phys.},  {\bf 104}, 3338 (1996). 

E.G. Timoshenko, Yu.A. Kuznetsov, K.A. Dawson,    
{\it Phys. Rev. E},
  {\bf 53}, 3886 (1995). 

\bibitem{SPhT}  
E. Pitard, H. Orland   
{\it preprint SPhT 96/118}. 

\bibitem{EDW}  
S.F. Edwards, P. Singh,    
{\it J. Chem. Soc. Faraday Trans. }II,  {\bf 75}, 1001
(1979). 

\bibitem{Flory}
P. Flory, {\it Principles of Polymer Chemistry},
Cornell 
University Press, Ithaca, N.Y. (1971). 

\bibitem{desCloizeaux}
 J. des Cloizeaux, { \it J. Phys. (France)\/}, {\bf 31}, 715 (1970).

\bibitem{DOI}  
M. Doi, S.F. Edwards   
{\it The Theory of Polymer Dynamics}, Clarendon Press, Oxford
(1986). 

\bibitem{thirum}
D. Thirumalai, { \it J. Phys. (France)} I, {\bf 5}, 1457 (1995).

\bibitem{TIM6}  
E.G. Timoshenko, Yu.A. Kuznetsov, K.A. Dawson,    
{\it Phys. Rev. E,}
  {\bf 54}, 4071 (1996). 

\bibitem{Thirum_Ash_Bhat}
D. Thirumalai, V. Ashwin and J.K. Bhattacharjee, {\em
Phys. Rev. Lett.}, {\bf 77}, 5385 (1996).

\bibitem{Roan_Shakh}
J-R. Roann and E.I. Shakhnovich, {\em Phys. Rev} E, {\bf 54}, 5340 (1996).

\bibitem{Dominicis_Laine_Orland}
C. De Dominicis, H. Orland and F. Lain\'ee, {\em J. Phys. (France)},
{\bf 46}, L-463 (1985). 

\bibitem{Koper_Hilhorst}
G.J.M. Koper and H.J. Hilhorst, {\em Europhys. Lett.}, {\bf 3}, 1213
(1987).

\bibitem{Shakhnovitch_Gutin}
E.I. Shakhnovich and A.M. Gutin, {\em Europhys. Lett.}, {\bf 9}, 569 (1989).

\bibitem{Iori_Marinari_Parisi}
G. Iori, E. Marinari and G. Parisi, {\em Europhys. Lett.}, {\bf 25},
491 (1994).

\bibitem{Wolynes_Onuchic}
J.D. Bryngelson, J.N. Onuchic, N.D. Socci and P.G. Wolynes, {\em
Proteins Struct Funct. Genet.}, {\bf 21}, 167 (1995).

\bibitem{Bouchaud}
see the review by J.P. Bouchaud, L. Cugliandolo, J. Kurchan and
M. M\'ezard in this book.

\bibitem{chaperon}
R.J. Ellis {\em Nature}, {\bf 328} 378 (1987)

R.J. Ellis {\em Curr. Opin. Struct. Biol.}, {\bf 4 
}, 117-122 (1994).

R.B. Freedman in
{\em Protein Folding'} ,  
Creighton T.E. 1992, (editor), W.H. Freeman, New York

\bibitem{Todd}
M.J. Todd, G.H. Lorimer and D. Thirumalai,
{\em Proc. Natl. Acad. Sci.},  {\bf 94}, 4030 (1996).

\bibitem{Orland_Thirumalai}
H. Orland and D. Thirumalai, {\em J. Phys. (France)}, I, {\bf 7}, 553 (1997).

\bibitem{Shakh_Gut1}
E. Shakhnovich, G. Fadtzinov, A.M. Gutin and M. Karplus, {\em
Phys. Rev. Lett.}, {\bf 67}, 1665 (1991).



\bibitem{PGG}
P.G. de Gennes, {\em Scaling concepts in polymer physics},
Cornell University Press, Ithaca (1979). 
\bibitem{Kaf}
K. Freed, {\em  Renormalization group theory of macromolecules}
, Wiley, New York, (1987).
\bibitem{Neg_O}
J.W. Negele and H. Orland, {\em Quantum Many Particle Systems}
, Addison Wesley, Menlo Park (1987).
\bibitem{Edw}
S.F.  Edwards, {\em Proc. Phys. Soc. London}, {\bf 85}, 613 (1965).
\bibitem{Orla_dd}
H. Orland, C. Itzykson and C. De Dominicis,  {\em J. Phys.(France)}, {\bf 46}, 
L-353 (1985).
\bibitem{Ma_Sch}
M. W. Matsen and M. Schick, {\em Phys. Rev. Lett.}, {\bf 72}, 2660 (1994).
\end{thebibliography}
\end{document}